\documentclass[]{elsarticle}
\usepackage[T1]{fontenc}

\usepackage{lineno,hyperref}
\usepackage{amssymb}
\usepackage{amsmath}
\usepackage{mathtools}

\usepackage{subfig,graphicx}
\usepackage{tikz}
\newlength\imageheight

\DeclarePairedDelimiter\abs{\lvert}{\rvert}%
\DeclarePairedDelimiter\norm{\lVert}{\rVert}%

\makeatletter
\let\oldabs\abs
\def\abs{\@ifstar{\oldabs}{\oldabs*}}
\let\oldnorm\norm
\def\norm{\@ifstar{\oldnorm}{\oldnorm*}}
\makeatother

\modulolinenumbers[1]

\journal{Physica A: Statistical Mechanics and its Applications}

\biboptions{numbers,sort&compress}

\begin{document}

\makeatletter
\def\ps@pprintTitle{%
  \let\@oddhead\@empty
  \let\@evenhead\@empty
  \let\@oddfoot\@empty
  \let\@evenfoot\@oddfoot
}
\makeatother

\begin{frontmatter}

\title{Decision-making with distorted memory: \\
Escaping the trap of past experience \tnoteref{title}}


\author{Evangelos Mitsokapas}
\author{Rosemary J. Harris}

\address{School of Mathematical Sciences, Queen Mary University of London, London E1 4NS, United Kingdom \\Correspondence to e.mitsokapas@qmul.ac.uk}

\begin{abstract}

Snapshots of ``best'' (or ``worst'') experience are known to dominate human memory and may thus also have a significant effect on future behaviour. We consider here a model of repeated decision-making where, at every time step, an agent takes one of two choices with probabilities which are functions of the maximum utilities previously experienced. Depending on the utility distributions and the level of noise in the decision process, it is possible for an agent to become ``trapped'' in one of the choices on the basis of their early experiences. If the utility distributions for the two choices are different, then the agent may even become trapped in the choice which is objectively worse in the sense of expected long-term returns; crucially we extend earlier work to address this case. Using tools from statistical physics and extreme-value theory, we show that for exponential utilities there is an optimal value of noise which maximizes the expected returns in the long run. We also briefly discuss the behaviour for other utility distributions. 
\end{abstract}


\end{frontmatter}


\section{Introduction}
Humans often make decisions based on evaluation of previous similar experiences. However, it turns out that rather than remembering an episode as a whole, people subconsciously only recall a selection of snapshots from the past. One way to encode such complex memory is the ``peak-end rule'' psychological heuristic. The peak-end rule, introduced by Kahneman \textit{et al.} \cite{kahneman93_1}, suggests that the recollected pain or benefit derived from an event (the remembered utility of a socio-economic scenario) can be predicted with substantial accuracy using a weighted average between the peak experience, represented usually by the most intense moment, and the final parts of the episode. An important consequence of this is the effect of ``duration neglect'' \cite{kahneman93_2}; retrospective evaluations of experiences often do not appear to depend significantly on their duration but rather, on the extreme and final snapshots. Evidence for the peak-end rule has been found in a wide range of contexts with utilities representing, for example, measures of pain in medical procedures \cite{kahneman93_3}, pleasure in living short but happy lives \cite{diener} and satisfaction associated with material goods \cite{doamy}.

An important, but less well-studied, question is how the peak-end rule for memory might affect future decisions. A first approach to quantify this involved a discrete-choice toy model with decision probabilities characterized by the maximum value of a utility random variable from all the time-steps when the same choice was made in the past \cite{rosemary}. For this model it was shown that for an agent sampling from a pair of choices with identical utility distributions, and decision-making subject to a predetermined level of noise, the evolution depends crucially on the tails of the utility distribution, leading to three classes of behaviour. Most interesting is the exponential-tail class, where there exists a particular value of noise marking a transition between a regime where the agent becomes ``trapped'' in one of the choices (i.e., consistently samples only one of the utilities available in the long-time limit) and a regime where the agent samples both decisions. This is connected to more recent research, where analogous classes of behaviour, under different types of memory kernels, govern whether an agent freely explores the space of choices or becomes trapped by ``force of habit'' \cite{bouchaud}. Other examples of trapping, for at least long periods of time, can be found in the behavioural economics literature; for instance, a revisited version of Kirman's classic ant toy model \cite{kirman} showcases how a single ant choosing to look for different food supplies can set off a torrent of ants following it, leading eventually to the colony switching to that food supply \cite{bouchaud_ants}. Additionally, trapping effects appear in the field of opinion dynamics where they relate to the tendency of a population to segregate into different viewpoints. In particular, a voter model based on many agents each with the peak memory used in \cite{rosemary}, reveals an interesting dependence on the noise \cite{polish}.

The main focus of the present paper is to extend the single-agent work of \cite{rosemary} to the case where the utility distributions for the two choices, are different. This is important, because in this generalization the agent faces the risk of becoming trapped in the worse choice, in the sense of expected long-term returns. We here investigate how different levels of noise affect the decision-making process and we show that, for exponential distributions,  there is an optimal value of noise, such that the agent can always escape the trapping pitfall and make the choice that maximizes its returns. As in \cite{rosemary}, this work involves a single agent, so we can use a mapping between the discrete-time decision model and a random walker with a particular kind of memory. Note that non-Markovian processes are interesting in themselves from a statistical physics point of view; for examples of how different forms of memory affect the long-term behaviour see \cite{hanggi, volkov, metzler, cressoni, kenkre, serva}. Indeed, there is a link between a simplified version of our model and the well-known elephant random walk \cite{schutz}; as we explain later, both are mathematically P\'olya urns \cite{polya} with the distinction that in our case there is a non-linear dependence on previous steps. Other work exploring the connections between P\'olya urns and reinforcement mechanisms in economics can be found in \cite{Bouchaud2021}. In particular, such models help highlight the importance of ergodicity-breaking, which is related to the distinction between ensemble averages and averages over trajectories.

Our work is organized in the following way. In Section \ref{section2} we present the discrete-choice model set-up as a random walker, explain connections and differences to P\'olya urns and summarize the results of \cite{rosemary} for the homogeneous case. In Section \ref{heterogeneouscasesection} we turn to new investigations for the heterogeneous case, in particular, using a simplified model to study the long-term behaviour for utilities with exponential-tails. In Section \ref{section 4} we demonstrate the existence of a special value of noise such that the average returns are optimized in the long-term, confronting predictions for the simplified model with simulation results for the real model. Finally, in Section \ref{section 5}, we conclude with a broader perspective on other distributions as well as a discussion on implications and open problems.

\section{Model set-up and extreme-value approximation}
\label{section2}
\subsection{Decision-making as a random walk}

In order to describe a decision model of two choices, we first consider a one-dimensional, discrete-time, random walker which at each time step $t$ moves either one step to the left or one step to the right. We denote the total number of right and left steps up to time $t$ by $X^+_t$ and $X^-_t$ respectively, such that $ X^+_t + X^-_t = t$. In addition, a trajectory time average can be defined as $V^{\pm}_t = X^{\pm}_t/t$, meaning $V^{+}_t + V^{-}_t = 1$. Consequently, at each step, position and average ``velocity'' are determined as $X_t = X_t^+ - X^-_t$ and $V_t = V_t^+ - V_t^-$, with the useful relation $V_t^{\pm} = (1 \pm V_t)/2$.

As mentioned in the introduction, we are interested in exploring how past experiences impact future choices. Thus, we here encode the history of experiences by assigning a (positive) utility random variable $U_i$ for each step $i$, with $1 \leq i \leq t$. The utilities are independent random variables, drawn from cumulative distribution functions $F^{\pm}(u)$ with right steps characterized by $F^+$ and left steps by $F^-$, respectively. Formally we have 
	\begin{equation}
	\label{fuprop}
	\mathrm{Pr} \left( U_i < u |X_{i} - X_{i-1} = \pm 1 \right) = F^{\pm}(u).
	\end{equation}
Note that in general the two choices can be heterogeneous, meaning $F^+ (u)\neq F^-(u)$; in particular, this means that one of the alternatives might be objectively ``better'', in the sense that it yields a higher average utility.

Within Kahneman's heuristic ``peak-end'' rule framework, the agent does not in fact remember the whole history of utilities, but instead remembers only the peak and the end. We focus here on the peak effect by keeping track of the maximum value of $U_i$ seen for all left steps up to the current time and separately the maximum for all right steps up to the current time. This gives rise to new history-dependent random variables $\hat{U_t}^{\pm}$, defined as
	\begin{equation}
	\label{max}
	\hat{U_t}^{\pm}= \max_{1 \leq i \leq t}\{ U_i: X_{i} - X_{i-1} = \pm 1 \}.
	\end{equation}
As detailed below, we explore in this paper how such a distortion of memory affects future decisions, particularly in the heterogeneous case.\footnote{For the rest of our analysis we choose to suppress the time subscripts (also on some other quantities), where interpretation should be obvious from context.}

To include the effect of $\hat{U}^+, \hat{U}^-$ on the dynamics we choose right and left transition probabilities for the random walker according to the ``logarithm of the odds'' (logit) convention:
	\begin{equation}
	\label{transitionprop}
	P^{+} \left(\hat{U}^{\pm}; T \right) = \frac{e^{\hat{U}^{+}/T}}{e^{\hat{U}^{+}/T}+e^{\hat{U}^{-}/T}}, \quad P^{-} \left( \hat{U}^{\pm}; T \right) = 		\frac{e^{\hat{U}^{-}/T}}{e^{\hat{U}^{+}/T}+e^{\hat{U}^{-}/T}}.
	\end{equation}
Here $T \in \mathbb{R}$ is a positive parameter, so the walker is more likely to move in the direction of the larger of $\hat{U}^+$ and $\hat{U}^-$. To ensure that both choices are equally likely in the beginning and the first step is always ``fair'', we impose initial conditions of $(\hat{U}^+_0, \hat{U}^-_0) = (0,0)$. It is worth noticing that the system, despite being non-Markovian in the position space, remains Markovian within a state space where $\hat{U}^{\pm}$ are included.

Note that the parameter $T$ represents the level of noise in the decision-making. Indeed, in the case of large values of $T$, the transition probabilities are $P^{\pm} \approx 1/2$, i.e., the two choices are approximately equally likely and the effect of memory is negligible. On the other hand, in the case of small values of $T$, there is a strong bias in the direction corresponding to the larger of $\hat{U}^+$ or $\hat{U}^-$. Furthermore, for a given trajectory, $\hat{U}^{\pm}$ increases monotonically with the number of steps in the corresponding direction. Hence, if there is a bias in say the right direction, the walker is more likely to move to the right and $\hat{U}^+$ is likely to increase further, thus strengthening the bias. This can lead to trapping by ``force of habit'' even with $F^+(u) = F^-(u)$, i.e., homogeneous choices~\cite{rosemary, bouchaud}. However, the interplay between noise and memory is expected to be more significant in the heterogeneous case $F^+(u) \neq F^-(u)$, since then there is the possibility that the agent can be trapped in the ``worse'' choice.

To enable a more precise analysis of the long-time behaviour, we introduce in the next Section a simplified model which can be treated as a generalized P\'olya urn but still captures the main features of the full model.

\subsection{Connection to P\'olya urns}

The dynamics of the walker depend on the history of $\hat{U}^{\pm}$. As hinted above, the values of $\hat{U}^{\pm}$ are correlated with the number of left and right steps in the past. Such a behaviour is reminiscent of a P\'olya urn problem \cite{polya}, where the probability of selecting an object of a specific characteristic (e.g., colour) depends on the proportion of that characteristic chosen previously. In the standard elephant random walk \cite{schutz} the transition probabilities are linear functions of the proportion of right and left steps taken in the past; a generalization of interest is the case where the transition probability is a nonlinear function of the proportion, see e.g., \cite{mahmoud, kearney}. This model is known as a generalized P\'olya process. However, in our model $P^{\pm}$ are not deterministic functions of the velocities $V^{\pm}$, rather functions of $\hat{U}^{\pm}$ which are, in turn, correlated random variables depending on the number of left and right steps. Here, following \cite{rosemary}, we approximate $\hat{U}^{\pm}$ with the corresponding characteristic largest values, such that the dynamics becomes similar to a kind of generalized P\'olya urn problem. To be concrete, given distribution functions $F^{\pm}$, the characteristic largest value after $X^{\pm}$ trials is defined as the value of $u$ such that $F^{\pm} (u) = 1 - 1/X^{\pm}$ \cite{gumbel}. Approximating the extreme values $\hat{U}^{\pm}$ with these characteristic largest values implies that the transition probabilities \eqref{transitionprop} depend, at the next step, on the number of previous left or right steps as required for a generalized P\'olya urn problem. However, note that the resulting probabilities depend explicitly on the number of left and right steps $X^{\pm}$, rather than the fractions $V^{\pm}$. This trait, depending on the form of the characteristic largest value, may result in direct time-dependence in \eqref{transitionprop}.

Using the characteristic largest value framework clearly neglects some fluctuations but it allows analytical progress. 
In the remainder of this paper we refer to this approximation as the ``simplified model'' and aim to compare its predictions to simulation results for the ``real model''. Some care is required to match the initial conditions of the two models although this is expected to only affect short-time properties. In order to predict the long-time limiting behaviour of the walker we are particularly interested in the existence of fixed points and the nature of their stability.

\subsection{Long-time behaviour}

One might anticipate that the system will eventually converge, for large $t$, to a fixed-point state where transition probabilities remain constant from one step to the next. In particular, we are interested whether the agent continues to sample both decisions in the long-time limit (i.e., the walker moves in both directions) or becomes ``trapped'' in one of the choices (i.e., the walker moves asymptotically in only one of the directions). The former might imply a stable fixed point with non-zero values of  $V^+, V^-$. The latter would correspond to stable fixed points at $(V^+, V^-) = (0, 1)$ and $(V^+, V^-) = (1, 0)$.

To explore the approach to fixed points in the long-time limit of the simplified model, we assume that in the velocity space there is a concentration around typical trajectories. When the random variable $V$ takes a value $v$, the mean distance moved in the next step is given by the corresponding value of the difference $P^+ - P^-$, denoted by $\Delta(v, t)$. This observation indicates that typical trajectories can be characterized by the discrete mapping:
\begin{equation}
\label{delta}
v_{t+1} = \frac{t v_t + \Delta(v_t, t)}{t + 1},
\end{equation}
where $v_t$ is the (average) value the random variable $V$ takes at the time-step $t$. Here $\Delta(v_t, t)$ depends explicitly on the functional form and parameters of the cumulative distribution functions $F^{\pm}$, as well as the value of noise $T$ in the decision.\footnote{We assume throughout that $\Delta (v_t, t)$ is a monotonically increasing function of $v_t$, consistent with the idea of positive reinforcement.} In the case where $\Delta(v_t, t)$ is time independent, we can investigate the stability of any fixed points $v_*$, satisfying $\Delta(v_*, t) = v_*$, by looking at the slope of $\Delta(v_t, t)$ at the fixed point. Specifically, if $ \partial{\Delta(v_t, t)}/\partial{v_t} |_{v_t = v_*} < 1$, then ``small'' fluctuations above or below $v_*$ are characterized by $\Delta(v_t, t) < v_t$ or $\Delta(v_t, t) > v_t$ respectively, thus on average the velocity converges back to $v_*$, indicating a stable fixed point. Analogously, $ \partial{\Delta(v_t, t)}/\partial{v_t} |_{v_t = v_*} > 1$ indicates an unstable fixed point.

In the case where $\Delta(v_t, t)$ depends on $t$, we can still look for time-dependent solutions to the equation $\Delta(\tilde{v}(t), t) = \tilde{v}(t)$. These solutions can be interpreted as attracting or repelling trajectories. In what follows, we are particularly interested in the long-time behaviour of such time-dependent trajectories, as well as the way in which an agent converges towards attracting trajectories in the velocity phase-space. For both stable fixed points and attracting trajectories, the decay is expected to be a form of power law which, by construction of the model, cannot be faster than the Markovian case of $1/t$. Furthermore, as will become clear later on, the noise parameter $T$ can play an important role in controlling the stability of such trajectories.

\subsection{Homogeneous case}

For the reader's convenience, we summarize here some results for the homogeneous decision-making model \cite{rosemary}, where $F^+ = F^-$. Within the extreme-value theory framework ones finds three different scenarios for the simplified model, which are heuristically argued to represent three generic distribution families. In particular, we explain the behaviour of the dynamics when the utility distributions are characterized by bounded, power-law or exponential tails. Note that, in the homogeneous case, $v_* = 0$ constitutes a fixed point for all three scenarios and the nature of its stability determines the long-time behaviour of the agent.

For a bounded distribution (i.e., one with a finite support), $v_* = 0$ is found to be always stable for long enough times and it is expected that the system approaches a random walk with $P^+$, $P^-$ fixed and equal. Similarly, for a power-law distribution, at long times, $v_* = 0$ is found to be unstable. In fact, as $t \to \infty$, $\Delta(v) \to \text{sgn}(v)$, so $v_* = \pm 1$ are stable fixed points of the system.\footnote{In more general heavy-tailed distributions, where the tail decays slower than an exponential function but faster than a power-law the situation appears to be slightly more subtle with a rather slow convergence to the relevant long-time behaviour in that case [I. D. Ajjour, personal communication]; such discussions are of course related to the domains of attraction of the universal distributions in extreme-value theory \cite{DeHaan2006}, \cite{Embrechts1997}.} This corresponds to a ``trapping'' phenomenon, where the agent keeps making the same choice, for long times. For both bounded and power-law distributions, we emphasize that the limiting behaviour of the agent, within the extreme-value approximation, is independent of the level of noise. This is consistent with numerical results for the real model.

We now give more details for the interesting case of exponential-tail distributions which will be most relevant for the remainder of the paper. For a pure exponential distribution with parameter $\lambda$, we calculate
\begin{equation}
\label{expdeltahom}
\Delta(v_t, t) = \text{tanh}\left( \frac{1}{2 \lambda T} \ln \frac{1 + v_t}{1 -v_t} \right).
\end{equation}
Significantly, here $\Delta(v_t, t)$ has no explicit dependence on time $t$, positioning the system in the standard class of generalized P\'olya urn models for which many mathematical results are available \cite{mahmoud, kearney}. Evaluating the slope of \eqref{expdeltahom} at $v_*=0$, yields an expression controlled by noise $T$ and the respective distribution parameter:
\begin{equation}
 \left. \frac{\partial{\Delta(v_t, t)}}{\partial{v_t}}\right|_{v_t = 0} = \frac{1}{\lambda T}.
\end{equation}
According to the stability condition, this fixed point is unstable at all times, for values of noise $T< 1/\lambda$, while it is stable for values such that $T > 1/\lambda$. Hence for $T < 1/\lambda$ we have trapping behaviour controlled by the stable fixed points $v_* = \pm 1$, similar to the power-law case. Correspondingly, one predicts that for $T > 1/\lambda$, the dynamics of the system approach those of a symmetric random walk with the velocity distribution converging to a $\delta$-distribution in the long-time limit.\footnote{In fact there is a further transition within this regime between super-diffusive fluctuations for $T < 2/\lambda$ and diffusive fluctuations for $T>2/\lambda$ \cite{rosemary, kearney}.} This prediction for the simplified model is confirmed by simulation results. The general picture of a transition from low to high-noise regimes is also observed in the real model but simulations suggest the existence of a finite-width velocity distribution around $v_*=0$ in the high-noise regime, even in the long-time limit. This finite width could be related to the weak-ergodicity breaking seen in \cite{bouchaud}; a detailed study of the correlations in the different noise regimes is a topic for further investigation. This, in turn, indicates the properties of the model depend on the full distribution of $\hat{U}^{\pm}$, rather than just the corresponding characteristic largest value approximation. For other exponentially tailed distributions one may find a slope with a weak (logarithmic) time-dependence; in the Gaussian case, for example, this is such that $v^* = 0$ is always stable in the long-time limit. However, for large but finite times, one still observes the generic transition between low and high-noise regimes.

We now turn to the heterogeneous case, where the situation is more complicated as $v_t = 0$ is no longer a fixed point.

\section{Heterogeneous decisions with exponential utilities}
\label{heterogeneouscasesection}
\subsection{Typical behaviour}

The trapping observed in the homogeneous case is expected to also be a feature in the heterogeneous case ($F^+ \neq F^-$), which is potentially more interesting, since the agent could be trapped in the wrong choice, i.e., the one with the lower average utility. We focus here on exponential utility distributions, inspired by the stability transition seen in the homogeneous case.

The full model is set up with parameters $\lambda^+ >0$ and $\lambda^- >0$ (with $\lambda^+ \neq \lambda^-$) such that the utilities used are drawn from cumulative distribution functions $F^{\pm}(u) = 1 - e^{-\lambda^{\pm} u	}$, for steps to the right or to the left respectively, with $(\hat{U}^+_0, \hat{U}^-_0) = (0,0)$ as before. In the corresponding simplified model, we start from the same initial conditions but for $X^+>0, X^->0$, we substitute the characteristic largest values, i.e., we approximate $ \hat{U}^+ = \ln X^+/\lambda^+$ and $ \hat{U}^- = \ln X^-/\lambda^-$. This returns the simplified-model approximation of $P^+$ and $P^-$, in terms of the average velocities $V^{\pm}$:
\begin{align}
\label{probexp}
	&P^+ = \frac{\left( V^{+} \right)^{1/\left( \lambda^{ + } T \right)}}{ \left(  V^+ \right)^{1/\left( \lambda^{+} T \right)}+t^{(\lambda^+-\lambda^-)/(\lambda^+ \lambda^- T)} \left( V^- \right)^{1/\left( \lambda^{-} T \right)}},  \\
	&P^- = \frac{\left( V^{-} \right)^{1/\left( \lambda^{ - } T \right)}}{ t^{(\lambda^--\lambda^+)/(\lambda^+ \lambda^- T)} \left(  V^+ \right)^{1/\left( \lambda^{+} T \right)}+\left( V^- \right)^{1/\left( \lambda^{-} T \right)}}.
	\end{align}
Contrary to the homogeneous case, probabilities here are explicitly dependent on time which leads to a time-dependence in the mean displacement of the next step:
\begin{equation}
\label{expdeltahet}
\Delta \left(v_t, t \right) = \text{tanh}\left[ \frac{1}{2T}  \ln{ \left( \left( \frac{2}{t}  \right)^{\frac{\lambda^+ - \lambda^-}{\lambda^+ \lambda^-}} \frac{(1+v_t)^{1/ \lambda^+ }}{ (1-v_t)^{1/ \lambda^-}  } \right)} \right].
\end{equation}
We emphasize here that $\Delta$ depends on the parameters $\lambda^{\pm}$ and $T$, which will play an important role in the upcoming analysis.

\begin{figure}[ht]%
\settoheight{\imageheight}{\includegraphics[height=5cm]{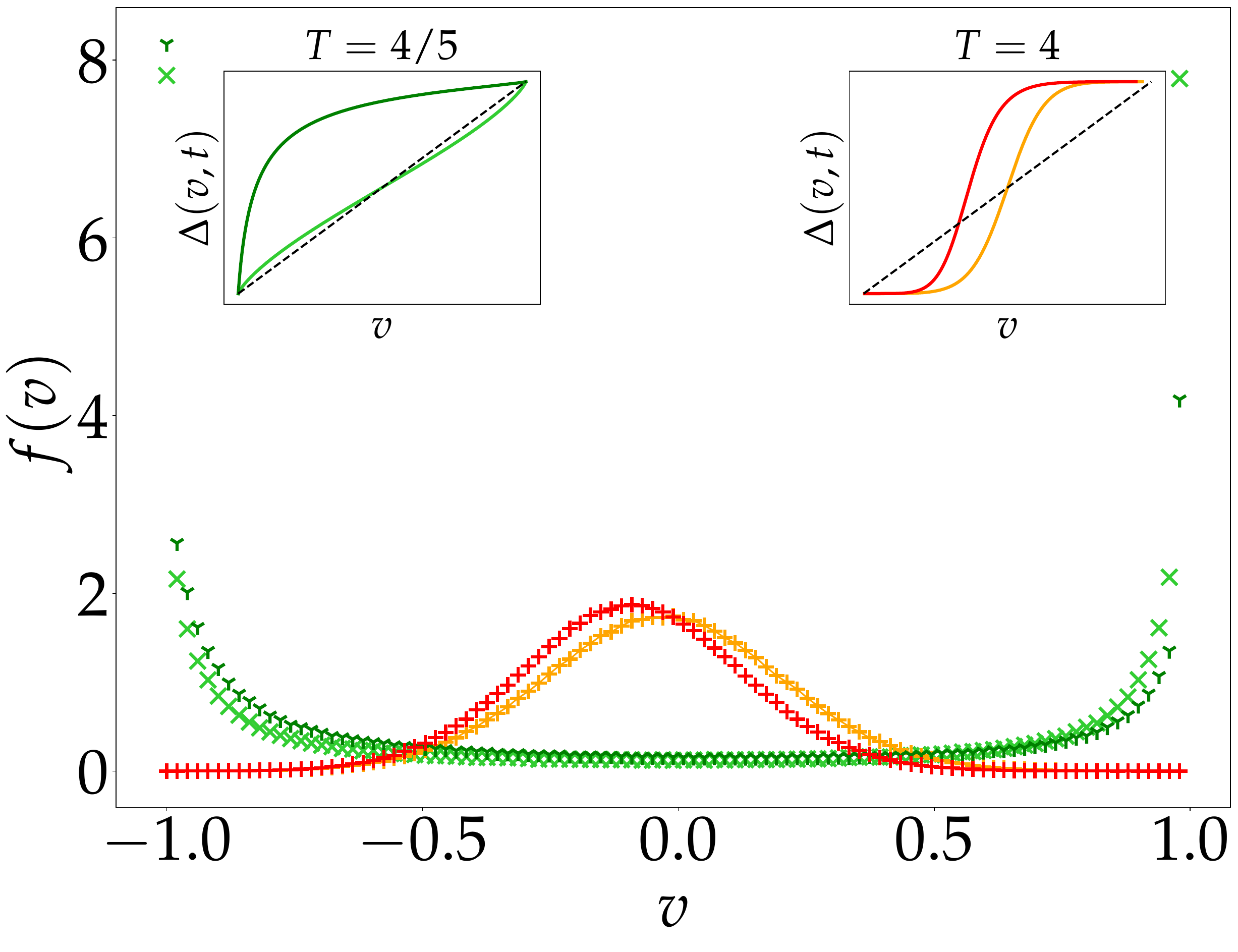}}
    \centering
\subfloat[\label{fig:img1}]{\includegraphics[height=5cm]{low_vs_high_multi_a.pdf}}%
\hfil
\subfloat[\label{fig:img2}]{\tikz\node[minimum height=\imageheight]{\includegraphics[height=3.5cm]{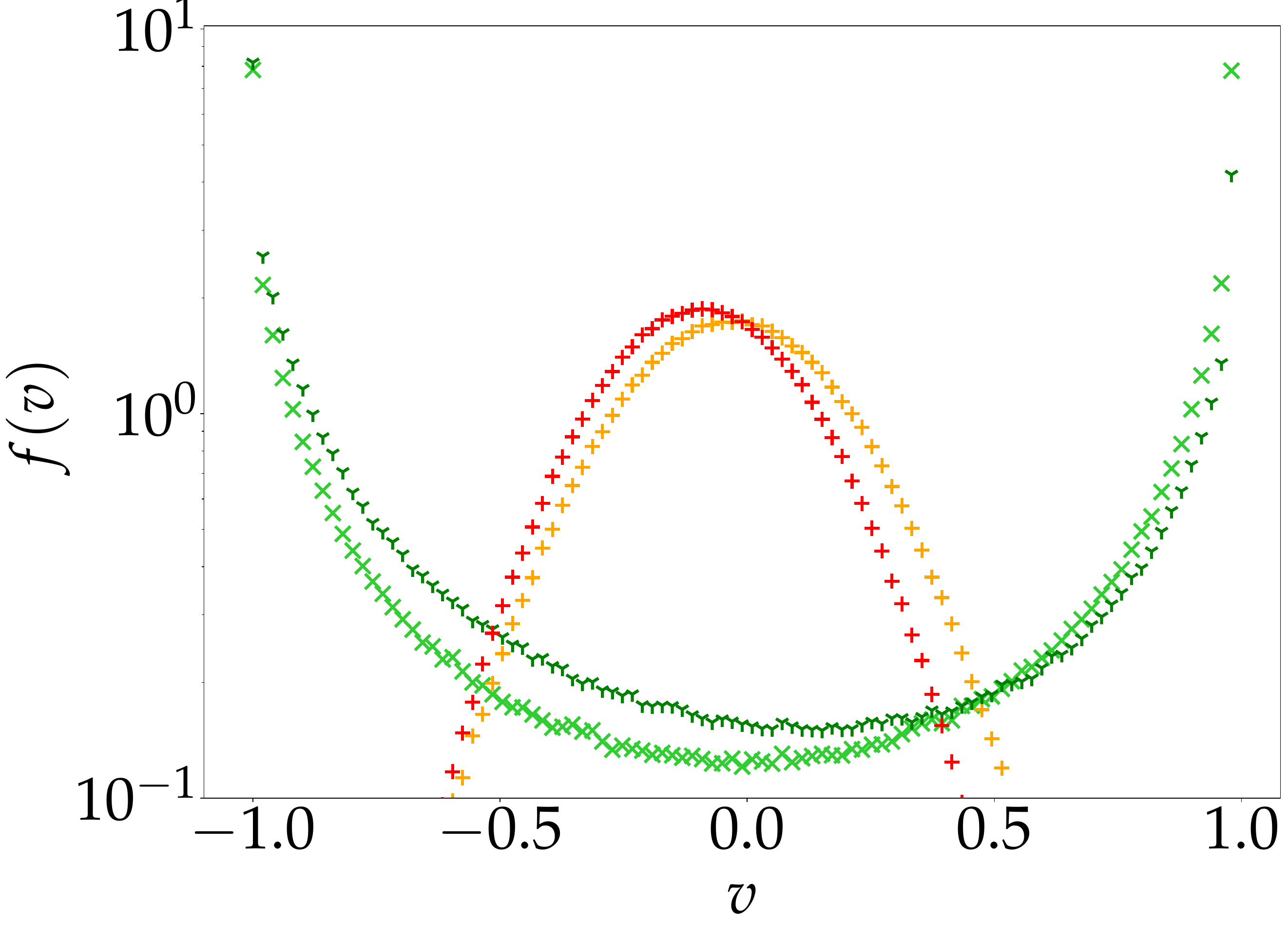}}; }%
    \caption{Empirical histograms of final velocity $V = V^+ - V^-$ for random walkers with dependence on the peak values of exponential utility distributions (calculated from $10^6$ trajectories, each running up to $t= 100$ time-steps). The asymmetric heterogeneous case (red and dark green points, $\lambda^+ = 6/5, \lambda^- = 1$) is compared with the symmetric homogeneous case (light green and orange points, $\lambda^+ = \lambda^- = 1$). In both cases there is a clear distinction between noise $T = 4/5$ (light/dark green) and $T = 4$ (red/orange). In the former case the asymmetry leads to a difference in relative height of the peaks at $v_* = -1$ and $v_* = +1$; the log-scale plot on the right further highlights the differences for intermediate velocities. The inset cobweb diagrams show the solutions to \eqref{fixedpoint}.}%
\label{fig:heterogeneroushist}%
    \end{figure}

Recall that the fixed points of the system, are time-independent solutions to
\begin{equation}
\label{fixedpoint}
\Delta \left(v_t, t \right) = v_t.
\end{equation}
 We observe here that $v_* = \pm 1$ are still fixed points of the system, but, significantly, $v_t = 0$ no longer satisfies the transcendental equation \eqref{fixedpoint}, except for the special case of $t = 2$. Instead we find time-dependent solutions  $\tilde{v}_t$ and their long-time behaviour is the focus of our attention. Similarly to the homogeneous case, we can determine their stability by calculating the slope $\partial{\Delta \left(v_t, t \right)}/\partial{v_t}$ evaluated at $\tilde{v}_t$. Random walker trajectories are expected to be typically found close to stable, time-dependent solutions $\tilde{v}_t$. This is supported by simulations, see the velocity histograms in Fig.~\ref{fig:heterogeneroushist}.
Similar to the homogeneous case, a distinction in behaviour according to the choice of noise can be observed. In the remaining part of this Section, we examine this in more detail by studying the evolution of time-dependent solutions to \eqref{fixedpoint}.

Without loss of generality we choose $\lambda^+>\lambda^-$ which implies that $v_* = -1$ can be characterized as the optimal choice of the agent, as it corresponds to a greater average utility, than $v_* = + 1$. We remark that there are two special values of noise, $T^+ = 1/\lambda^+$ and $T^- = 1/\lambda^-$, which yield closed-form solutions to \eqref{fixedpoint}: the trajectories $\tilde{v}^+_t = 1-2/t$ and $\tilde{v}^-_t = 2/t-1$, respectively. The corresponding slope of $\Delta$, evaluated at $\tilde{v}^+_t $ gives
\begin{equation}
\frac{\partial}{\partial{v_t}} \Delta\left(v_t, t \right)\big|_{\tilde{v} = \tilde{v}^+_t} = \frac{\lambda^- + \lambda^+ (t-1)}{\lambda^- t},
\end{equation}
and it is then a straightforward task to show that $\tilde{v}^+_t $ is an unstable trajectory, since $\lambda^- + \lambda^+(t-1) > \lambda^- t $, for all $t>1$. Analogously, evaluating the slope of $\Delta$ at $\tilde{v}^-_t$, returns
\begin{equation}
\frac{\partial}{\partial{v_t}} \Delta \left(v_t, t\right) \big|_{\tilde{v} = \tilde{v}^-_t} = \frac{\lambda^+ + \lambda^- (t-1)}{\lambda^+ t}.
\end{equation}
and hence $\tilde{v}^-_t$ is a stable trajectory.

It turns out that these special values of noise $T^+$ and $T^-$ mark transitions between three distinct noise regimes which we refer to as low noise, intermediate noise and high noise. To illuminate the behaviour within these regimes we generally need to resort to numerical methods to investigate the time-dependent solutions and their stability. We stress that the fixed points $v_{*} = \pm 1$ are independent of the level of noise and therefore exist in all three regimes.

\subsection{Three noise regimes}

\subsubsection{Low noise}

\begin{figure}
    \centering
{{\includegraphics[width=\textwidth]{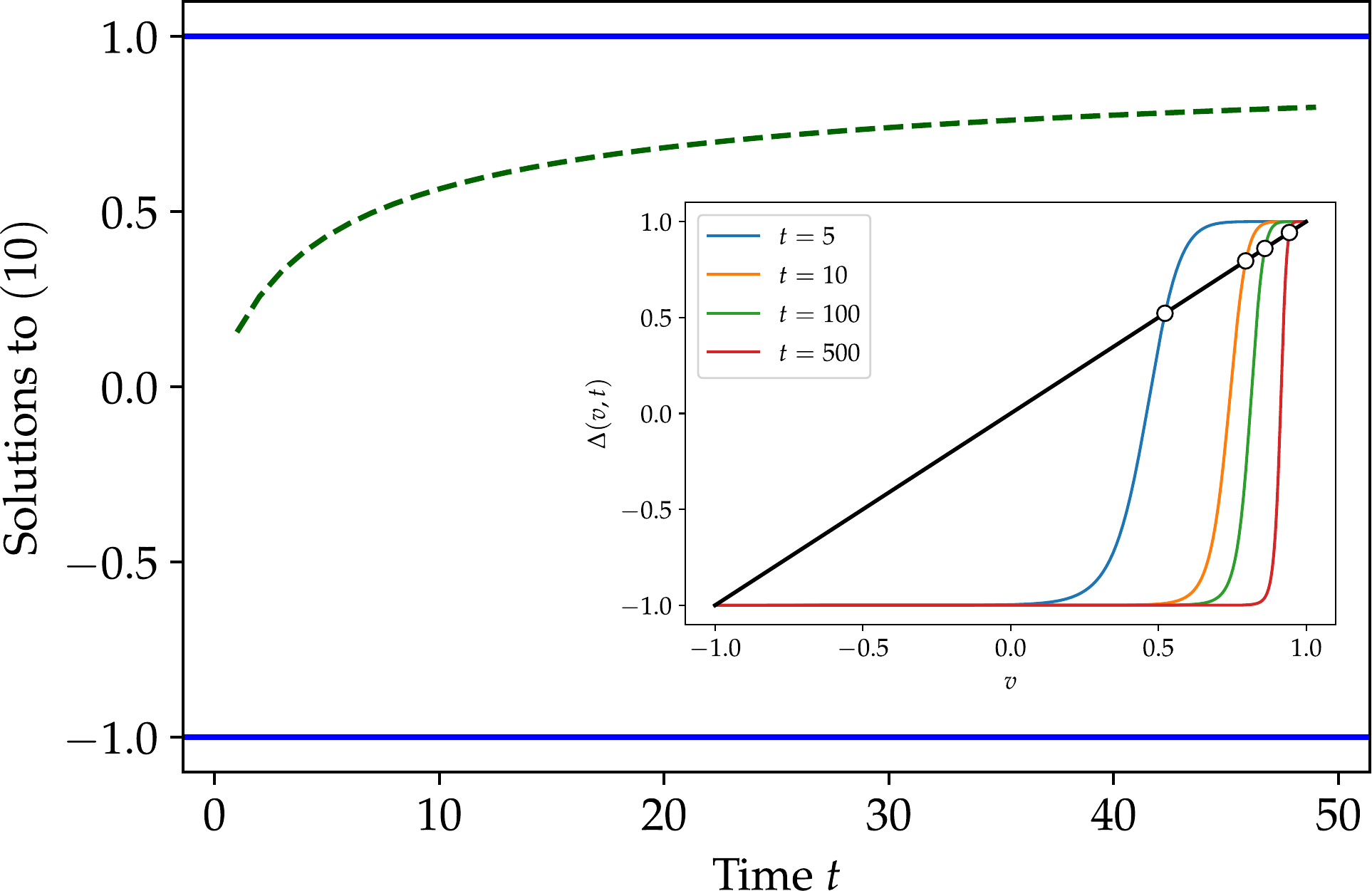}}}%
    \caption{Plot of the solutions to $\Delta \left(v_t, t \right) = v_t$ and inset cobweb diagram for the low-noise region, $T < T^+$. Numerics yield a time-dependent unstable $\tilde{v}_t$ (dashed green), along with stable $v_{*} = \pm 1$ (solid blue). Results are for $(\lambda^+, \lambda^-) = (2,1)$ and $T = 1/8$.}
    \label{fig:fplessthanTplus}%
\end{figure}

We first turn to the low-noise regime, which corresponds to $T < T^+ = 1/\lambda^+$. Here, the behaviour is controlled by stable fixed points $v_{*} = \pm 1$, as well as an unstable trajectory $\tilde{v}_t$. Numerics in Fig.~\ref{fig:fplessthanTplus} indicate that $\tilde{v}_t \to + 1$, in the long-time limit.
Indeed, we notice that $\tilde{v}_t$ appears to decay to $v_* = + 1$ rather slowly. An analytical solution to \eqref{fixedpoint}, obtained for another special case with parameter values $(\lambda^+, \lambda^-) = (2,1)$ and~$T = 1/4$ suggests this is a power-law decay. This picture indicates that for low noise, just as for the homogeneous case, the agent becomes trapped in one of the two choices (corresponding to the stable fixed points $v^* = \pm 1$). In other words, in the long-time limit, it is found in a ``frozen'' state in which the same decision is always made. In contrast to the homogeneous case, the asymmetry of the present problem means there would generally be an unequal likelihood of ending up in $v^* = +1$ or $v^* =  -1$. We later explore how these relative probabilities, and hence the ensemble-averaged utility, depend on the level of noise.

\subsubsection{Intermediate noise}

\begin{figure}
    \centering
{{\includegraphics[width=\textwidth]{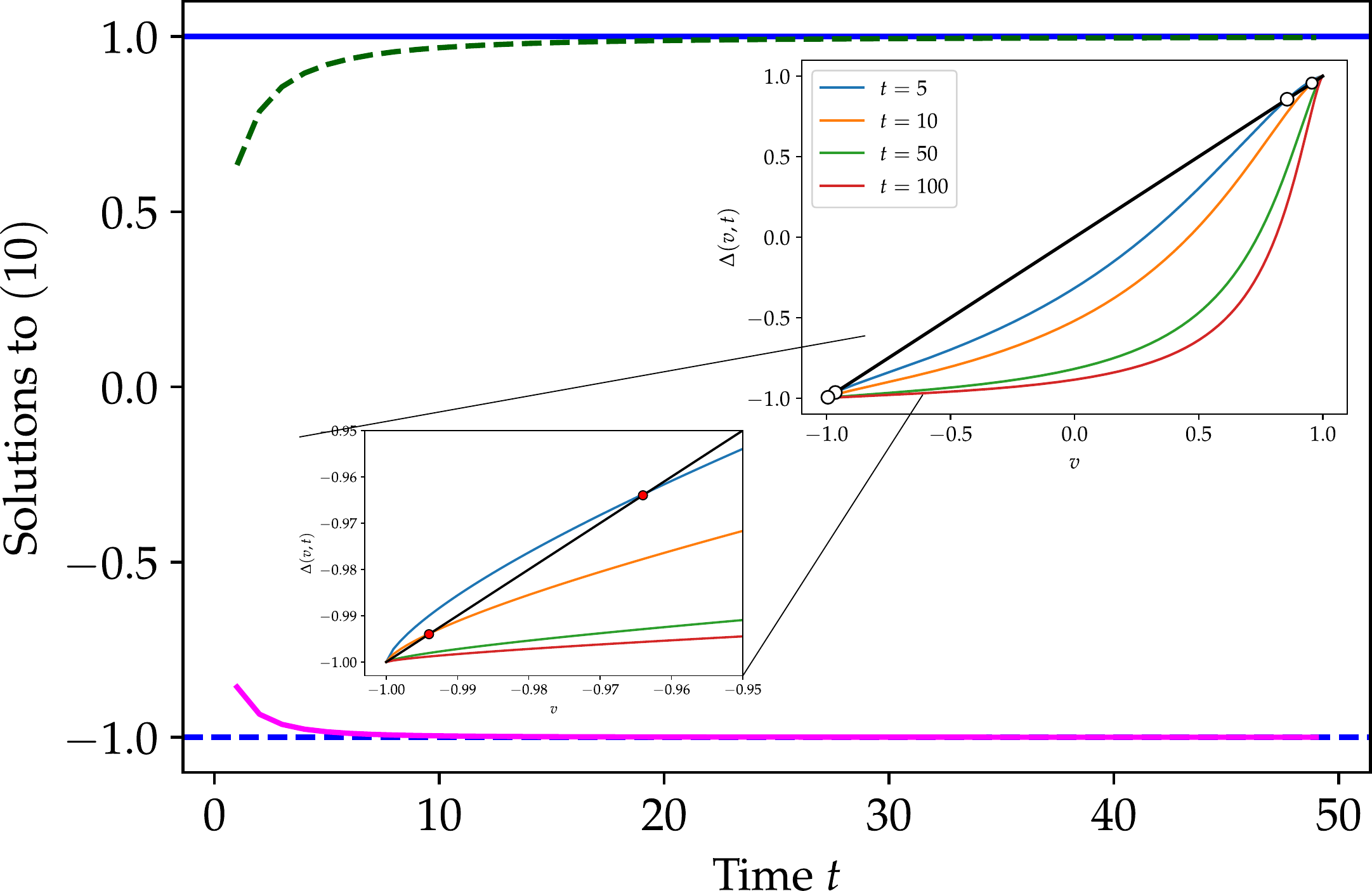} }}%
    \caption{Same as Fig.~\ref{fig:fplessthanTplus} but for intermediate-noise case ($T = 7/10$), with an additional blow-up of the cobweb diagram near $v_* = -1$. Here there are unstable (dashed green) and stable (solid magenta) trajectories, $\tilde{v}^{\pm}_t$, which co-exist alongside the stable (solid blue) and unstable (dashed blue) fixed points at $v_* = \pm 1$.}
    \label{fig:fpbetweenTplusTminus}%
\end{figure}

A different picture emerges in the intermediate-noise regime, $T^+ < T < T^-$. Here, there is an interplay between co-existing trajectories satisfying \eqref{fixedpoint}, $\tilde{v}^+_t$ and $\tilde{v}^-_t$, converging to  $+1$ and $-1$, respectively, in the long-time limit. The evolution of these trajectories is plotted in figure~\ref{fig:fpbetweenTplusTminus}.
We find that $\tilde{v}^{-}_t$ is stable, whereas $\tilde{v}^{+}_t$ is unstable; correspondingly the fixed points $v_* = -1$ and $v_* = +1$ are unstable and stable respectively. Numerical results suggest both the intermediate trajectories decay towards the fixed points in a power-law like manner, with $\tilde{v}^{-}_t$ converging at a different rate to $\tilde{v}^{+}_t$, as shown in the inset cobweb diagram in figure~\ref{fig:fpbetweenTplusTminus}. For short times the agent may be found close to the stable fixed point $v^* = +1$. However, the crucial question, to be explored later, is whether this trapping persists in the long-time limit or the agent can escape to $v^* = -1$. Note that this intermediate-noise regime has no parallel in the homogeneous case, where $T^+ = T^-$.

\subsubsection{High noise}

\begin{figure}
    \centering
{{\includegraphics[width=\textwidth]{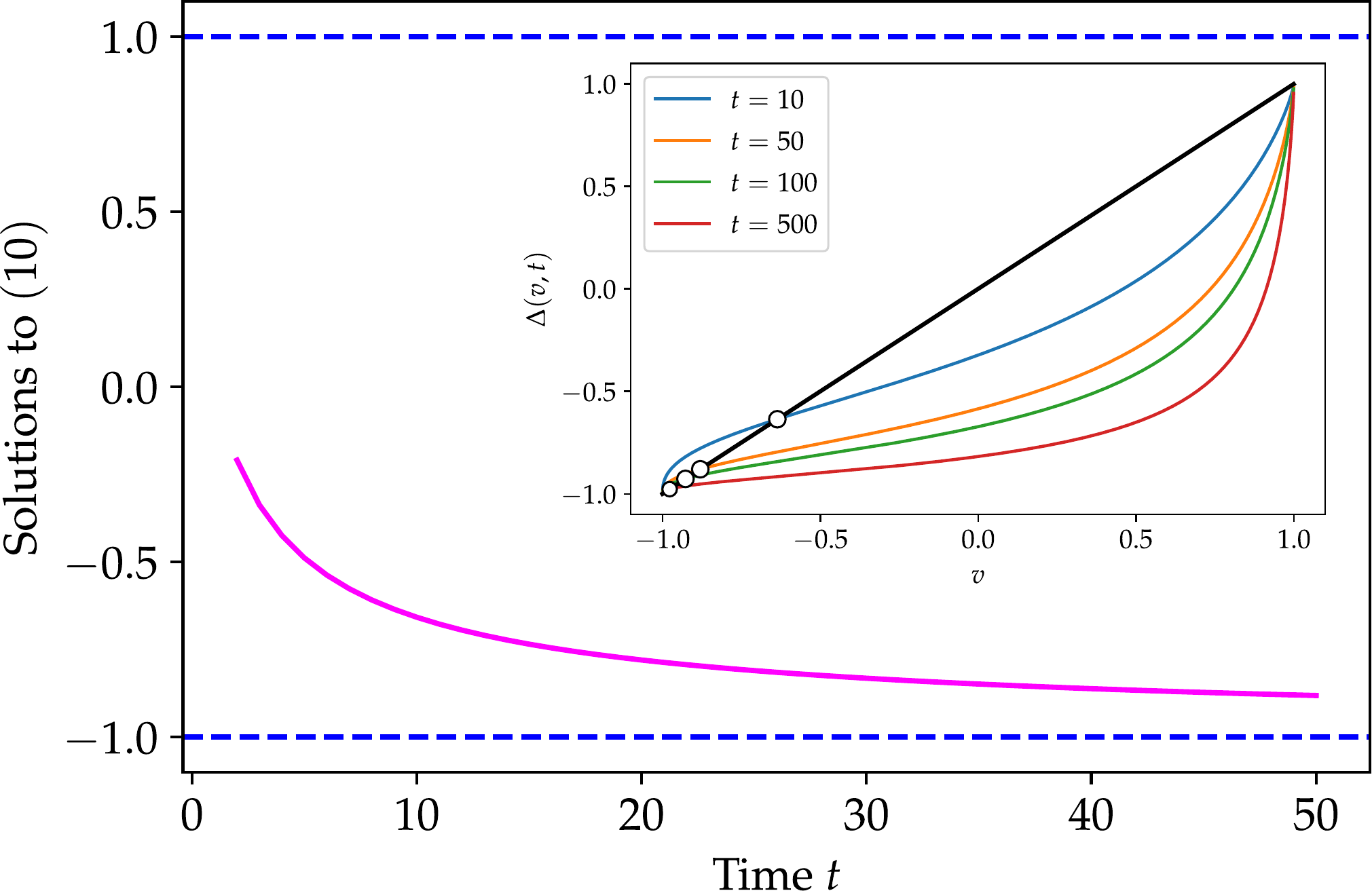} }}%
    \caption{Same as Fig.~\ref{fig:fplessthanTplus} but for high-noise case ($T = 6/5$). Fixed points $v_* = \pm 1$ (dashed blue) are unstable while $\tilde{v}_t^{-}$ (solid magenta) is stable.}
    \label{fig:fpaboveTminus}%
\end{figure}

In the high-noise regime, $T > T^-$, the behaviour of the system is controlled by unstable fixed points $v_* = \pm 1$ and a single stable trajectory $\tilde{v}_t$. The corresponding cobweb diagram in Fig.~\ref{fig:fpaboveTminus} reveals $\tilde{v}_t \to -1$ in the long-time limit. 
However, numerical calculations indicate the corresponding convergence rate is significantly smaller than in the intermediate-noise regime. This means that for large, but finite, times the walk of the agent is typically characterized by velocities fluctuating about the stable $\tilde{v}_t$, still sampling both of the choices, but very slowly decaying towards $-1$. In the corresponding regime in the homogeneous case, $v_* = \pm 1$ are also unstable but significantly the intermediate trajectory is replaced by a stable fixed point at $0$. 

In the next Section we turn our attention to how the noise dependence of the long-time trajectory behaviour affects the ensemble average utility. Numerics indicate that the real and simplified model share crucial common features, so we continue to concentrate our analysis on the latter. 

\section{Optimization of expected returns}
\label{section 4}


\subsection{Expected utility and the dynamical landscape picture}

As became apparent in the stability analysis of the $\tilde{v}_t$ satisfying \eqref{fixedpoint}, it is possible that, for utilities drawn from two different exponential distributions, the agent becomes confined in the worse choice. We still consider, without loss of generality, $\lambda^+>\lambda^-$ such that the worse choice corresponds to right steps.

\begin{figure}
    \centering
  \subfloat[]{\includegraphics[width=0.5\textwidth]{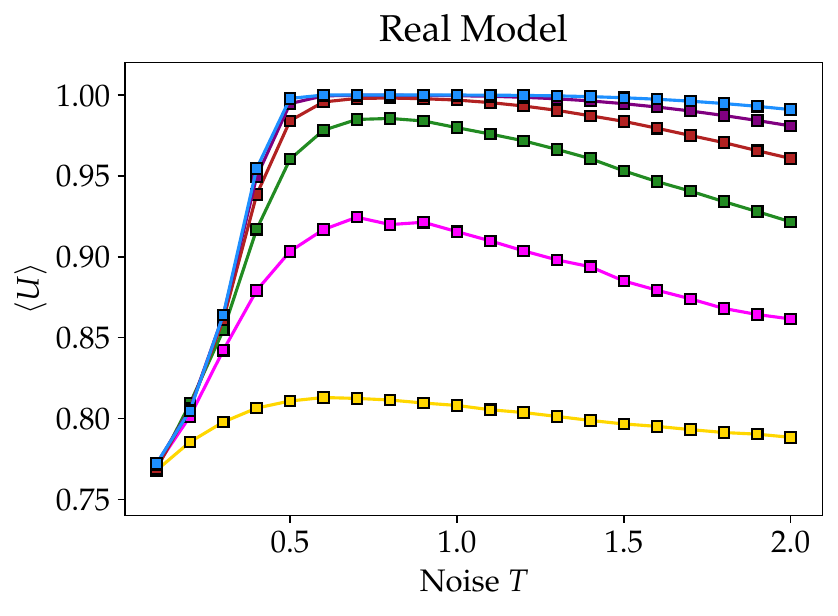}\label{fig:realmodel}}
  \subfloat[]{\includegraphics[width=0.5\textwidth]{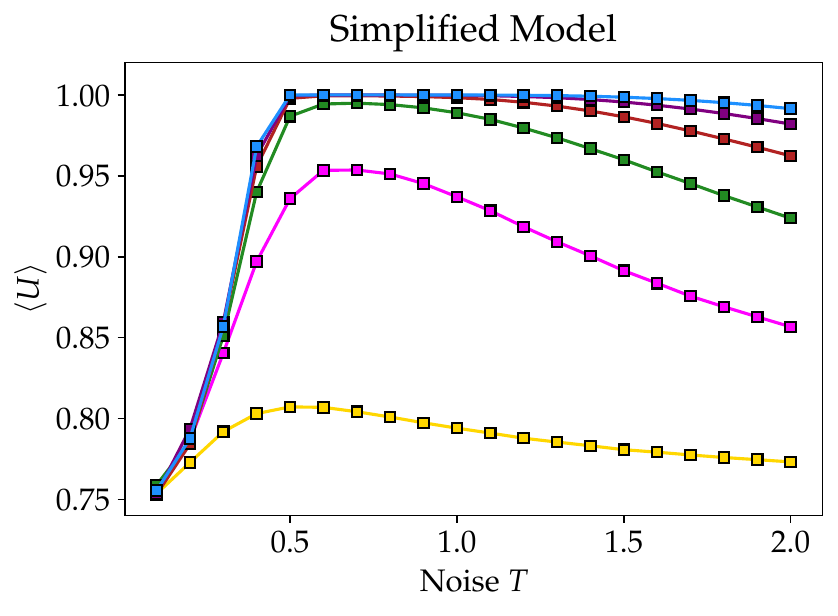}\label{fig:simplifiedmodel}}
    \caption{Comparison of agent average returns for the real model and its extreme-value  approximation. Averaging has been implemented over time as well over the ensemble of the simulation. Points show simulation results with averages over increasing numbers of time-steps; $10$ (yellow), $10^2$ (magenta), $10^3$ (green), $10^4$ (red), $10^5$ (purple) and $10^6$ (blue). Results are for $10^5$ trajectories in each case, with $(\lambda^+, \lambda^-) = (2,1)$.}
    \label{fig:comparisonaverageutility}%
\end{figure}

To quantify this phenomenon, we here study the average utility per agent, per time-step. In Fig.~\ref{fig:comparisonaverageutility}, we show a numerical estimation of this expected utility value $\langle U \rangle$ (averaged over the ensemble of agents) for both the real model and its simplified counterpart, for different values of noise $T$.
Interestingly, we find that for both models there appears to be a non-monotonic dependence of $\langle U \rangle$ on $T$, suggesting there is an optimal value of noise, leading to a peak value for $\langle U \rangle$. To investigate in more detail, we revisit the noise regimes defined earlier in Section \ref{heterogeneouscasesection} focusing on the simplified model, which as we can see is itself a reasonable approximation to the real model.

In analysing the different regimes and especially the transition between low-noise and intermediate-noise regimes, it is useful to consider a landscape picture, defined by the potential-like quantity associated with $\Delta \left(v_t, t \right)$, given as
\begin{align}
\label{potentialeq}
\Phi(v_t,t) = t^{-1}\int_{0}^{v_t} \left( u - \Delta \left(u, t \right) \right) \text{d}u.
\end{align}
We fix the potential to equal zero at $v_t = 0$. Solutions to \eqref{fixedpoint} in Section \ref{heterogeneouscasesection} are obviously equivalent to the integrand in \eqref{potentialeq} being zero. Correspondingly, the associated potential graph will have zero slope at the position of these solutions. In particular, unstable fixed points map to local maxima while stable fixed points map to local minima, i.e., the potential landscape shows ``hills'' and ``valleys'' respectively.
This potential function allows for a visualization of the time-dependent (unstable) peak the agent has to overcome in order to transition to a (stable) valley. 

\subsection{Three noise regimes revisited}
\label{revisitednoiseregimes}

\subsubsection{Low noise}

In the low-noise regime, as $T \to 0$ the value of the expected utility appears to be invariant of the choice of total time $t$. Indeed, for small values of noise, even for longer times, each agent almost always makes the same choice as its initial step which, recall, is equally likely to be left or right. This means that approximately half of the ensemble will be at $v_* = +1$, leaving the other half at $v_* = -1$, so $\langle U \rangle = (1 / \lambda^+ + 1 / \lambda^-)/2 $. Memory for $T \approx 0$ has a strong effect on the dynamics, leading to the phenomenon of ergodicity-breaking with two steady states corresponding to the two stable fixed points. Recall that given $\lambda^+ > \lambda^-$, $v_* = -1$ constitutes the optimal state for the walker. Figure \ref{fig:comparisonaverageutility} then shows, for both models, that as noise $T$ is increased (but still $T<1/\lambda^+$), the agent is more likely to escape towards the optimal steady state, leading to an increase of the average return. A semi-analytical approximation for the expected returns in the low-noise region can be obtained building on results for first-passage times in P\'olya urns \cite{kearney} (see Fig.~\ref{fig:approximationlownoise} with more details in the Appendix).
\begin{figure}
    \centering
{{\includegraphics[width=\textwidth]{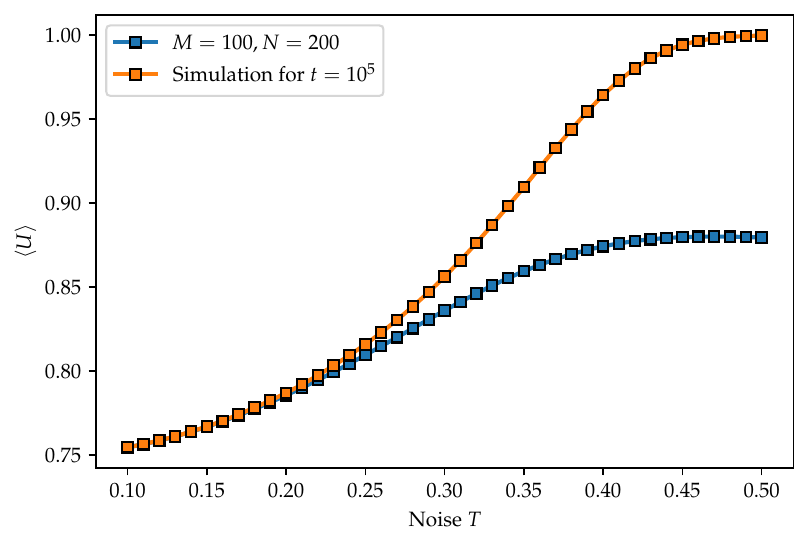} }}%
    \caption{Plot of the $\langle U \rangle $ from the simulation for the low-noise regime (orange) against the approximation (blue) derived from consideration of first-passage times in P\'olya urns. By construction, the approximation is better for small values of noise. For larger $T$, we expect the approximation to improve somewhat as the parameters $M$ and $N$ are increased (see Appendix).}
    \label{fig:approximationlownoise}%
\end{figure}

We now turn our attention to the dynamical behaviour close to $T  = 1/\lambda^+$, where it appears there is a peak of $\langle U \rangle$, at least for finite times. Indeed, it becomes clear from the ergodicity-breaking picture that the theoretical maximum utility $1/\lambda^-$ would be attained when all of the agents who became initially trapped in the wrong choice, were able to escape and become trapped in the optimal choice. Our simplified model simulation results show that as time increases the height of the peak increases (corresponding to more of the agents escaping towards the optimal steady state) and for long times the upper bound becomes saturated: $\langle U \rangle$ obtains its maximum value at $T = 1/\lambda^+$. For the real model an analogous phenomenon can be observed; here the time-dependence is more complicated, but for long times saturation occurs at $T  \approx 1/ \lambda^+$.

\begin{figure}
    \centering
  \subfloat[]{\includegraphics[width=0.5\textwidth]{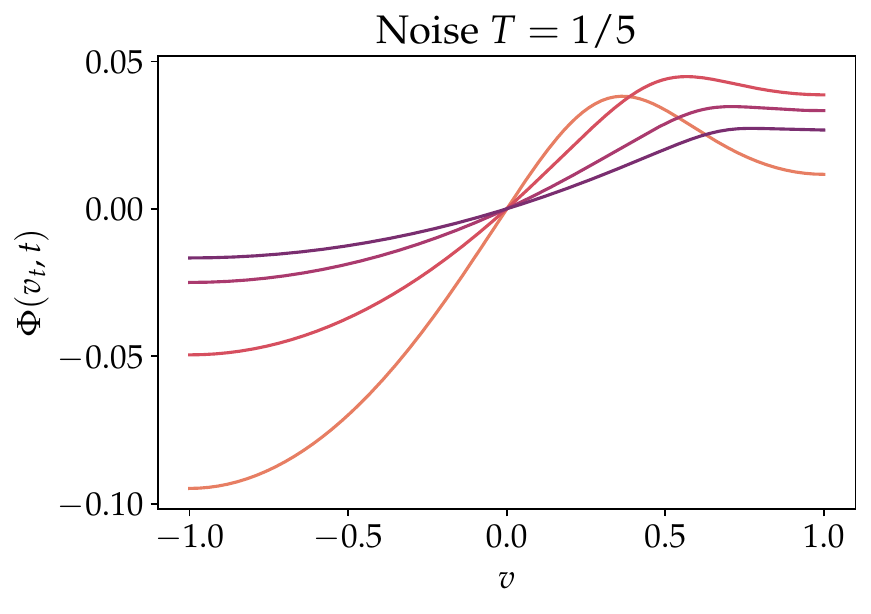}\label{fig:potentialnoise02}}
  \subfloat[]{\includegraphics[width=0.5\textwidth]{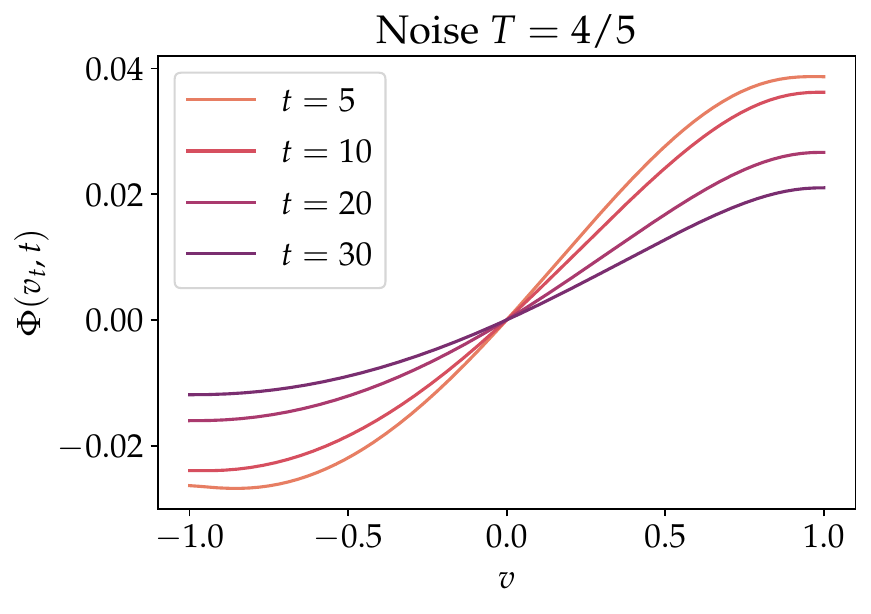}\label{fig:potentialnoise08}}
    \caption{Dynamical behaviour of the potential landscape at fixed values of $T$ in the low and intermediate-noise regimes for increasing times (light to dark). Legend is the same for both plots (a) and (b). Parameter choices $(\lambda^+, \lambda^-) = (2,1)$.}
    \label{fig:landscapepotential}%
\end{figure}

To explain this saturation, we consider the potential landscape in Fig.~\ref{fig:landscapepotential}(a).
For values of noise $T<1/\lambda^+$, the relative height of the peak to the valley bottom gradually becomes smaller as time $t$ increases. Despite that, in this low-noise regime, the agent can still become trapped in $v_* = +1$. A heuristic argument for this can be made by considering the time-dependence of the peak position compared to the rate at which the agent can move through the potential landscape. Specifically, we can consider what happens if the agent moves only to the right for the first $(t-1)$ time-steps such that the average velocity is $v_{t-1} = +1$, i.e., the agent can be found at the valley bottom corresponding to the fixed point $v_* = + 1$. To investigate whether the agent can indeed escape this frozen state, let us now assume that in the next time step $t$, it takes a rare (but not impossible due to fluctuations) step to the left, so the average velocity becomes $ v_{t} = 1- 2/t$.
Significantly, however, this average velocity is always greater than the velocity of the (unstable) trajectory $\tilde{v}_t$, that satisfies \eqref{fixedpoint}. In other words, a single step to the left is not enough for the agent to overcome the unstable hill and such a rare fluctuation is typically followed by a relaxation, back towards the valley bottom at $v_* = +1$. Furthermore, the relative effect of a single step to the left (compared to the position of the peak) decreases with time, such that escape becomes more and more improbable.

As $T$ is increased in the low-noise regime, fluctuations away from the typical trajectories obviously increase. At the special value $T = 1/\lambda^+$, the average velocity from a single step to the left, $v_t = 1 - 2/t$, coincides with the peak of the unstable hill. In other words, a single step to the left is often enough for the agent to escape and in the long time-limit the whole ensemble manages to overcome the hill and become trapped in the valley bottom, corresponding to the optimal choice $v_* =  - 1$. This leads to the saturation of the upper bound of $\langle U \rangle$. 

\subsubsection{Intermediate noise}

Moving to the intermediate-noise regime, $1/\lambda^+ < T < 1/\lambda^-$, the potential landscape picture differs only slightly (see Fig.~\ref{fig:landscapepotential}), with the valley at $v_* = +1$ still present and a single step to the left again typically enough for the agent to overcome the unstable hill. However, for finite times, $v_* = -1$ is no longer a valley but an unstable hill and the agent reaches instead the intermediate valley, given by the stable trajectory $\tilde{v}_t$. In fact, $\tilde{v}_t$ itself approaches $v_* = -1$ in the long-time limit, such that $\langle U \rangle$ is maximized and the upper bound again becomes saturated. We argue that the speed of the upper bound saturation is controlled not by the speed of convergence of $\tilde{v}_t^{}$ to $v_* = -1$ (which is relatively fast, as seen in Fig.~\ref{fig:fpbetweenTplusTminus}) but rather by  the speed the agent moves down the potential landscape, after overcoming the unstable hill. In particular, as we increase the level of noise beyond the special value of $T= 1/ \lambda^+$, we expect fluctuations against the slope to play a more significant role, meaning it takes longer for the agent to reach the valley bottom.

To make the above argument more concrete, we note that a typical velocity $v_{t}$ at time-step $t$ implies that on average at the next step, a fraction $(1 - \Delta(v_t, t))/2$ of the ensemble agents move to the left and a fraction $(1  + \Delta(v_t, t))/2$ to the right. Given that left steps have an expected utility $1/\lambda^-$ and right steps have $1/\lambda^+$, averaging over all time-steps we get
\begin{align}
\langle U \rangle  = \frac{1}{t}\sum_{\tau = 0}^{t-1} \left(  \frac{1 + \Delta(v_{\tau}, \tau)}{2} \times \frac{1}{\lambda^+} + \frac{1 - \Delta(v_{\tau}, \tau)}{2} \times   \frac{1}{\lambda^-} \right),
\end{align}
where $t$ is the total number of time-steps and $\Delta(0, 0) = 0$ by construction. This then simplifies to
\begin{equation}
\label{approximation}
\langle U \rangle = \left( \frac{\lambda^+ - \lambda^-}{2\lambda^+ \lambda^- t} \right) \sum_{\tau = 0}^{t-1} \left( 1- \Delta(v_{\tau}, \tau) \right)  + \frac{1}{\lambda^+}.
\end{equation}
In the long-time limit, we can replace the sum by an integral, resulting in the following approximation:
\begin{equation}
\label{approximationlogtime}
\langle U \rangle \approx \left( \frac{\lambda^+ - \lambda^-}{2\lambda^+ \lambda^- t} \right) \int_{}^{t} \left( 1- \Delta(v_{\tau}, \tau) \right)d\tau  + \frac{1}{\lambda^+}.
\end{equation}
Furthermore, we conjecture that in this intermediate-noise regime, a typical trajectory leads to a long-time behaviour $ \Delta(v_t, t) \sim -1 + \alpha_T t^{-1}$. This corresponds to logarithmic corrections to the standard Markovian $t^{-1}$ decay of $v_t$. Here, $\alpha_T$ is a coefficient which presumably depends on the value of noise $T$ and characterizes the speed at which the agent moves down the potential landscape, towards the (faster moving) valley bottom. Substituting this functional form for $\Delta(v_t, t)$ in \eqref{approximationlogtime} indicates that, in the long-time limit, the dependence on time $t$, for the ensemble expected utility, is captured by
\begin{equation}
\label{intermediatenoiseapprox}
\langle U \rangle \approx \frac{1}{\lambda^-}  - \frac{\alpha_T \ln{t} + \beta_T}{t},
\end{equation}
where $\beta_T$ is a constant depending on the short-time behaviour which cannot be predicted in this approach.  

\begin{figure}
    \centering
{{\includegraphics[width=\textwidth]{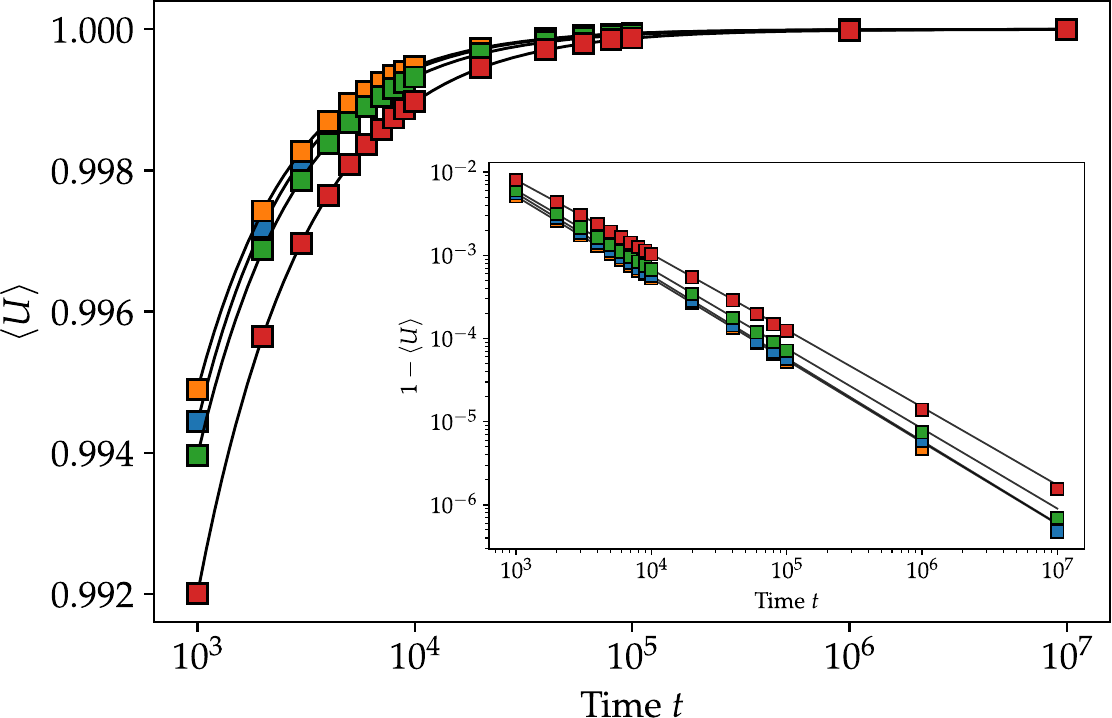} }}%
    \caption{Plot of the empirically determined $\langle U \rangle$ and the corresponding approximation \eqref{intermediatenoiseapprox}, against time $t$. Simulation is for $10^7$ trajectories with parameters $(\lambda^+, \lambda^-) = (2,1)$. Different marker colours represent different values of noise: blue for $T = 0.6$, orange for $T = 0.7$, green for $T = 0.8$ and red for $T = 0.9$. Associated fits, with solid black lines, are shown for each value of noise, with fit parameters: $ \left( \alpha_{0.6}, \beta_{0.6} \right) =  \left(0.08, 2.64 \right), \left( \alpha_{0.7}, \beta_{0.7} \right) =  \left(0.16, 2.28 \right), \left( \alpha_{0.8}, \beta_{0.8} \right) =  \left(0.65, 1.91 \right)$ and $\left( \alpha_{0.9}, \beta_{0.9} \right) =  \left( 2.01, 0.53 \right) $. Inset shows the same data on a log-log scale.}
    \label{fig:powerlaw}%
\end{figure}
The above long-time approximation appears to accurately capture the saturation of the expected value of $U$ with time (see fitted curves in Fig.~\ref{fig:powerlaw}). The numerics indicate that the $\alpha_T$ parameter increases with noise. In particular, for noise close to $1/ \lambda^+$,  the value of $\alpha_T$ is relatively small, hence the average utility approaches $1 / \lambda^-$ almost as a $t^{-1}$ power law, leading to a linear form in the log-log plot, in the inset of Fig.~\ref{fig:powerlaw}. However, as noise $T$ increases, the value of $\alpha_T$ increases as well and logarithmic corrections to the power law become significant (corresponding to a slowing in the speed at which a typical agent moves towards the valley bottom). As $T \to 1/ \lambda^-$, numerical results suggest that these logarithmic corrections diverge. At this point, the unstable hill merges with the stable $v_* = +1$. 

\subsubsection{High noise}

\begin{figure}
    \centering
  \subfloat[]{\includegraphics[width=0.45\textwidth]{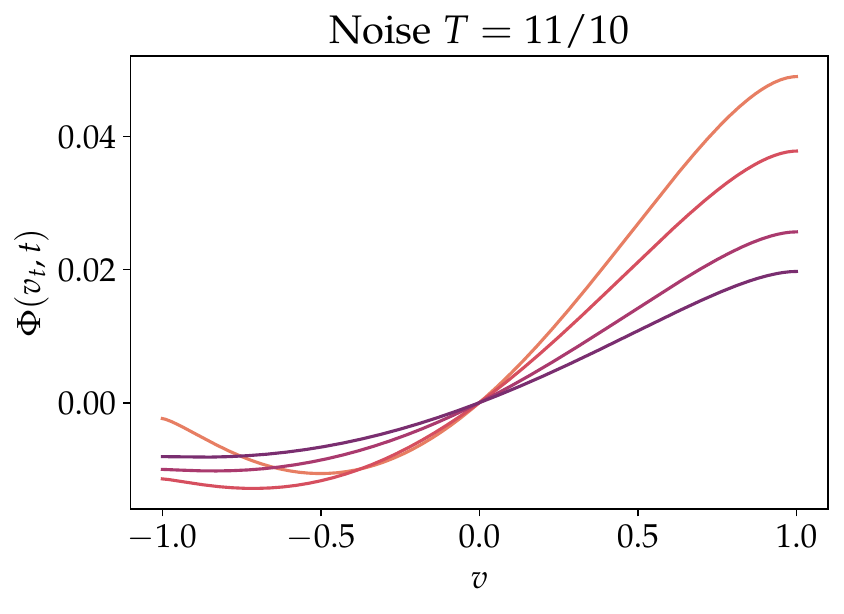}\label{fig:potentialnoise11}}
  \subfloat[]{\includegraphics[width=0.45\textwidth]{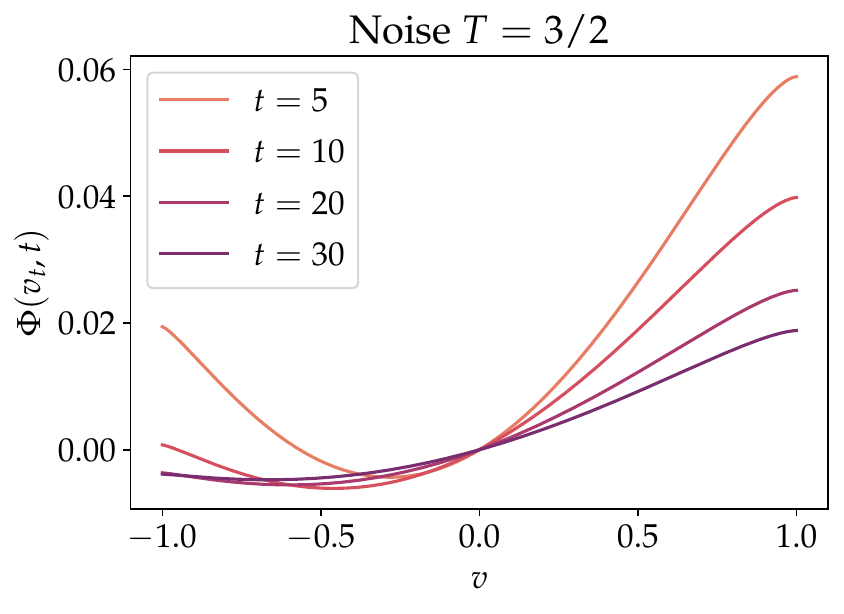}\label{fig:potentialnoise15}}
    \caption{Dynamical behaviour of the potential landscape at fixed values of $T$ in the high-noise regime. Time-steps and parameters same as Fig.~\ref{fig:landscapepotential}. Legend is the same for both plots (a) and (b).}
    \label{fig:landscapepotential_high}%
\end{figure}

In the high-noise regime, $T > 1/ \lambda^{-}$, the potential landscape differs drastically from that observed in the intermediate-noise regime (see Fig.~\ref{fig:landscapepotential_high}).
The fixed points $v_* = \pm 1$ are still present but, for finite times, they are both now unstable. There is only one intermediate solution to \eqref{fixedpoint}, corresponding to the stable valley bottom $\tilde{v}_t$. For a typical agent, a single step to the left still obviously suffices to move away from $v_* = +1$ and towards the valley bottom $\tilde{v}_t$. Similarly to the intermediate-noise regime, numerics suggest that $\tilde{v}_t$ converges to $v_* = -1$ in the long-time limit, maximizing $\langle U \rangle$ and thus leading to a saturation of the upper bound. To investigate how fast this saturation happens, we now consider the speed of convergence of the stable $\tilde{v}_t$ to $v_* = -1$. We hence look for solutions to \eqref{fixedpoint} of the form $\tilde{v}_t = -1 + \epsilon(t)$, anticipating that $\epsilon(t) \to 0$ as $t \to \infty$. This leads to the asymptotic form
\begin{equation}
\label{analyticalpowerlaw}
\tilde{v}_t  \sim -1 + 2t^{- \frac{\lambda^+ -  \lambda^-}{\lambda^- \left( T \lambda^+ - 1 \right)}}.
\end{equation}
Since $T > 1/ \lambda^-$, we crucially observe that the exponent of $t$ is greater than $-1$. This suggests that, throughout the high-noise regime, the approach of a typical agent to $v_* = -1$ is determined not by the decay to the valley bottom, but by the slower dynamics in \eqref{analyticalpowerlaw}, describing the movement of the valley bottom itself. As a consequence, the saturation of $\langle U \rangle$ is slower than in the intermediate-noise regime; effectively the system can be found in an adiabatic state where $\Delta(v_t, t)$ can be approximated by $\tilde{v}_t$ and substituting \eqref{analyticalpowerlaw} in \eqref{approximationlogtime} yields, to leading order,
\begin{equation}
    \label{approximationhighnoisetime}
    \langle U \rangle \approx \frac{1}{\lambda^-} +\frac{\left(\lambda^- - \lambda^+  \right)\left( T \lambda^+  - 1\right)}{\left(  \lambda^+ \right)^2 \left( T \lambda^-  - 1\right)} t ^{- \frac{\lambda^+ -  \lambda^-}{\lambda^- \left( T \lambda^+ - 1 \right)}}. 
\end{equation}

\begin{figure}
    \centering
{{\includegraphics[width=\textwidth]{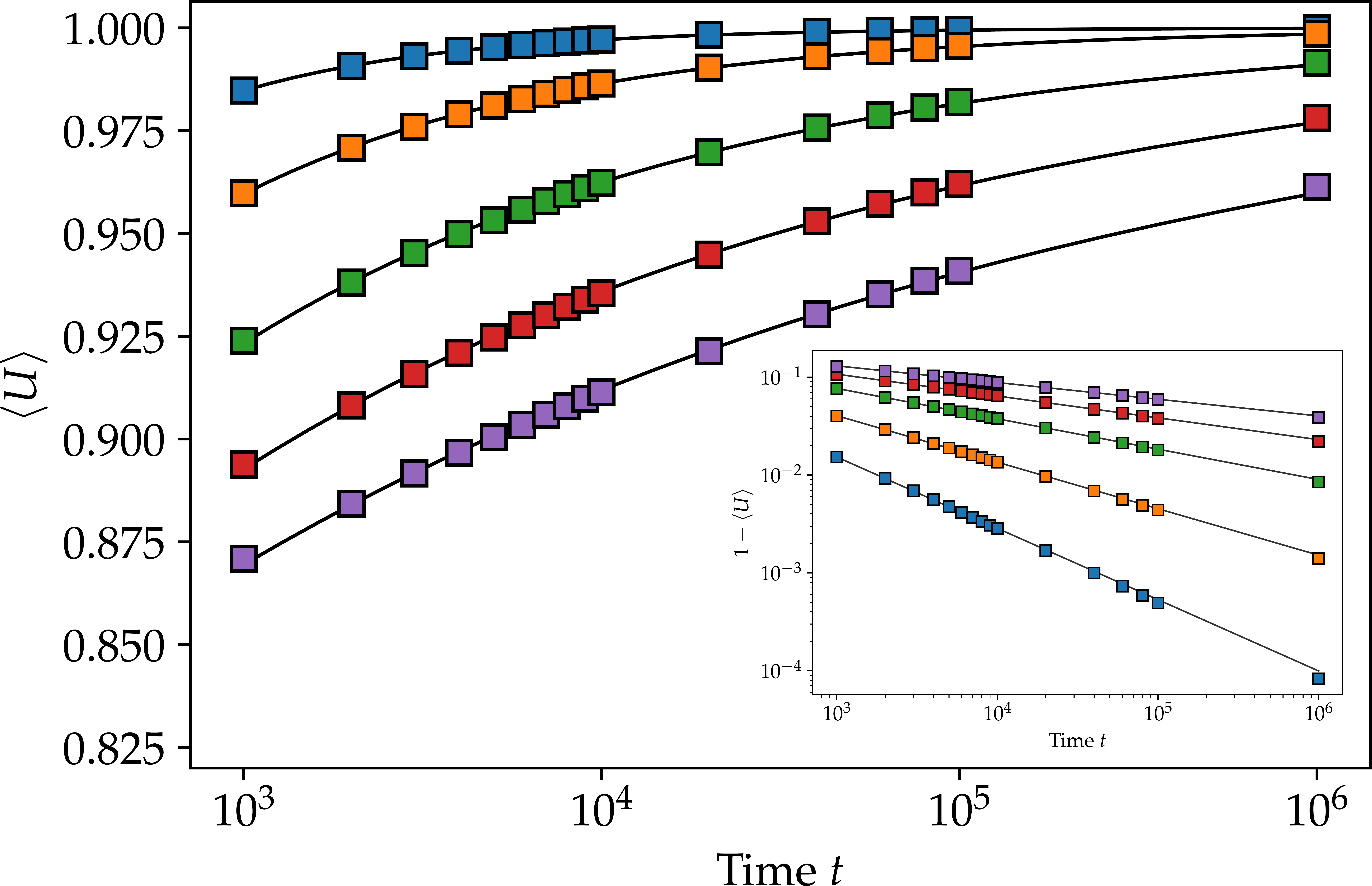} }}%
    \caption{Plot of the empirically determined $\langle U \rangle$ compared with the predicted functional form~$1/\lambda^- + \rho_T t^{\nu_T}$, against time $t$. Simulation parameters same as Fig.~\ref{fig:powerlaw}. Different marker colours represent different values of noise: blue for $T = 1.1$, orange for $T = 1.5$, green for $T = 2.0$, red for $T = 2.5$ and purple for $T = 3.0$. Associated fits with solid black lines, are shown for each value of noise, with fit parameters: $ \left( \rho_{1.1}, \nu_{1.1} \right) =  \left(-2.35, -0.73 \right), \left( \rho_{1.5}, \nu_{1.5} \right) =  \left(-1.08, -0.48 \right), \left( \rho_{2.0}, \nu_{2.0} \right) =  \left(-0.66, -0.31 \right)$, $\left( \rho_{2.5}, \nu_{2.5} \right) =  \left( -0.5, -0.22 \right)$ and $\left( \rho_{3.0}, \nu_{3.0} \right) =  \left( -0.42,  -0.17 \right)$. Inset shows the same data on a log-log scale.}
    \label{fig:powerlaw_high}%
\end{figure}

\begin{figure}
  \centering
{{\includegraphics[width=\textwidth]{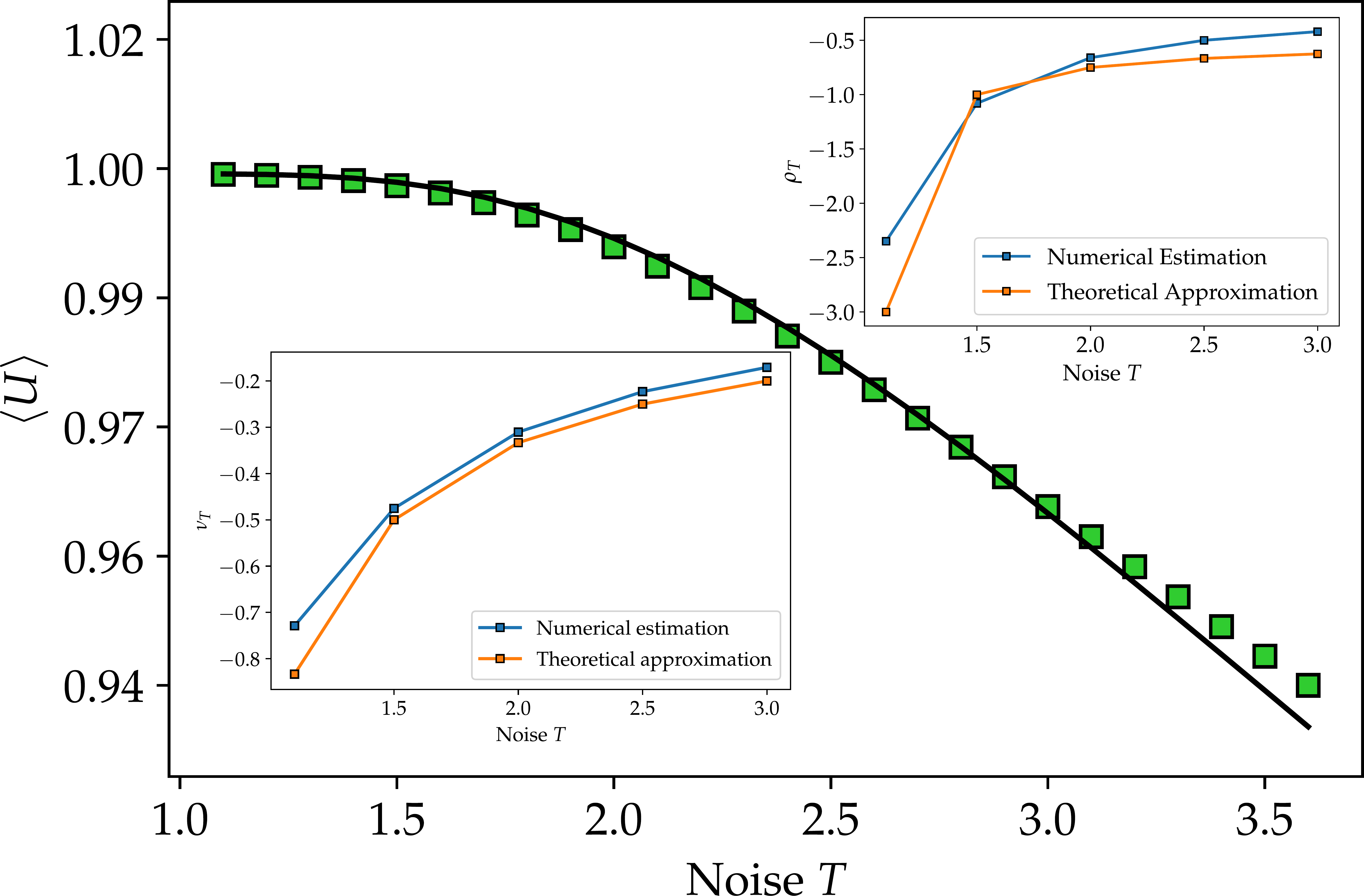} }}%
  \caption{Theoretical approximation (solid black line) and numerics (green) in the high-noise regime. Insets compare the exponent $\nu_T$ (bottom-left) and the prefactor $\rho_T$ (top-right) from numerical fitting with the numerical values from \eqref{approximationhighnoisetime}. Parameters as in Fig.~\ref{fig:comparisonaverageutility}.}
  \label{fig:exponents_and_approx}%
\end{figure}

Similar to the intermediate-noise regime we confront our predictions with empirical results for the simplified model, here via fitting a functional form ~$1/\lambda^- + \rho_T t^{\nu_T}$, see Fig.\ref{fig:powerlaw_high}. In this case we can also directly compare the numerics to the theoretical parameters. As seen in Fig.~\ref{fig:exponents_and_approx}, in a large part of the high-noise regime  \eqref{approximationhighnoisetime} provides a good long-time approximation to the numerics. This observation is supported by close agreement of fitted parameters from the numerics and the corresponding theoretical approximation; both the power-law exponent and the prefactor are well predicted by \eqref{approximationhighnoisetime}, as shown in the insets of Fig.~\ref{fig:exponents_and_approx}. The prediction of the exponent is less good for $T \approx 1/ \lambda^-$ where the time-scale of the valley bottom dynamics is close to $ t^{-1}$, hence sub-leading corrections play a more important role for finite times. Also, for very large values of noise, the power-law approximation is found to be less good for finite times. Indeed, we anticipate the existence of two non-commuting limits, regarding the values of noise $T$ and time $t$. In the case where $t \to \infty$ before $T \to \infty$, $\langle U \rangle$ will always converge to its maximum value; in practice, this means that for any finite value of noise, a typical agent will eventually maximize their expected returns (however slowly). In the case where $T \to \infty$ before $t \to  \infty$, $\langle U \rangle$ will always converge towards its minimum value, $\left(\lambda^+ + \lambda^-\right)/\left(2 \lambda^+ \lambda^- \right)$; this corresponds to the fact that for infinite values of noise, the memory does not play any role at all, as the agent always has an equal chance of making either a left or a right step. 

In closing this section, we note that although the detailed analysis has been for the simplified model, we expect the qualitative features to hold for the real model as well (compare the main features in Fig.~\ref{fig:comparisonaverageutility}).

\section{Discussion}
\label{section 5}
In this paper, we extended previous work \cite{rosemary} on a model of decision-making with peak-memory (inspired by Kahneman's heuristic ``peak-end'' rule) to consider the heterogeneous case, i.e., with asymmetric choices. Specifically, the model can viewed as a one-dimensional, discrete-time random walker (agent) assigned a utility random variable $U_i$ at each time step, which is sampled from a different distribution for left and right steps (corresponding to two different choices). The transition probabilities are defined to depend on the maximum values of $U_i$ for left and right steps as well as the level of noise $T$ in the decision-making process of the agent. We used the characteristic largest value of extreme-value theory to cast the problem as a generalized P\'olya urn model; this simplified version of the actual peak-memory model appears to accurately capture the dynamics while allowing more theoretical progress. In particular, we examine the stability of the fixed points of the system in the velocity phase-space. This is especially interesting in the case of heterogeneous choices where one of the two choices is preferable in the sense of having a higher average utility but early experiences could still lead to the agent becoming trapped in the choice with lower average utility. The probability of escaping such a trapping state can obviously depend on the value of noise in the decision.

The tails of the utility distribution also play a crucial role. Indeed, in the homogeneous case it was previously found \cite{rosemary}  that there are three classes of behaviour, derived from extreme-value theory. In particular, for exponential tails the corresponding simplified model shows (from a statistical physics point of view) a phase transition as a function of noise $T$ and this crossover behaviour is also evident in the real model. In this spirit, we made the focal point of our investigations here the case where left and right steps are both sampled from exponentially-tailed distributions but with two different parameter values, $\lambda^-$ for the left steps and $\lambda^+$ for the right steps (setting $\lambda^+ > \lambda^-$, without loss of generality). We looked particularly at the time-averaged expected returns of an ensemble of agents and, significantly, for finite times we found there exists a peak in these expected returns, for a specific value of noise $T$ (approximately $1/\lambda^+$). This peak appears to exist in both the real and simplified models but to gain theoretical understanding we concentrated on the simplified model. As time increases, in this model, the height of the peak rapidly becomes saturated at $1/\lambda^-$ which corresponds to all agents becoming trapped in the choice with the higher average utility. In fact, for longer times this saturation also occurs for values of noise greater than $1/\lambda^+$ with the asymptotic convergence displaying two different power-law regimes. We were able to explain this, and predict the power-law exponents as a function of the distribution parameters, via a dynamical potential landscape picture which helps to visualize the stability of fixed points and trajectories.
\begin{figure}
  \centering
  \subfloat[]{\includegraphics[width=0.5\textwidth]{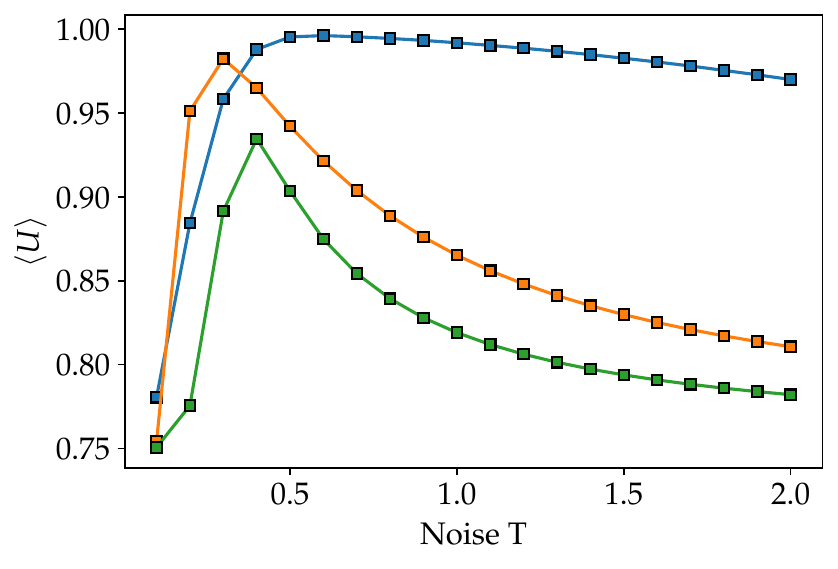}\label{fig:multipledist10_3}}
  \subfloat[]{\includegraphics[width=0.5\textwidth]{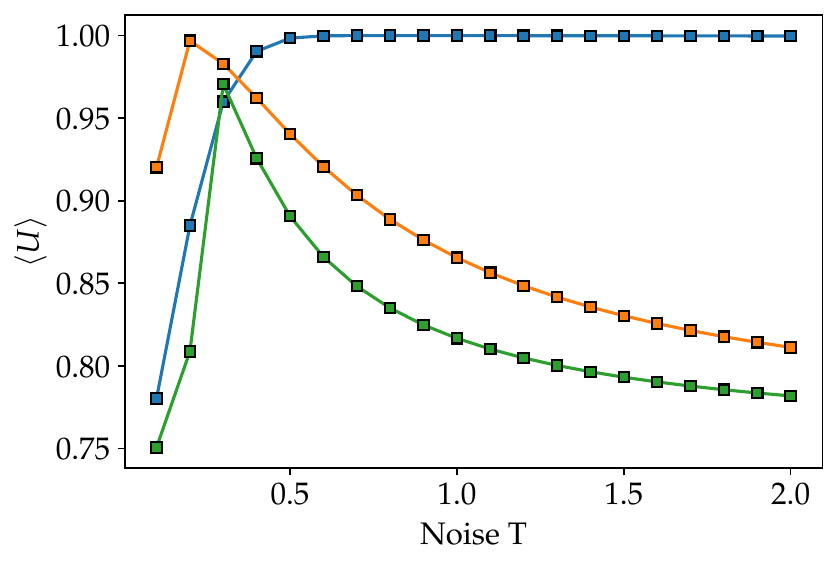}\label{fig:multipledist10_5}}
  \caption[]%
  {Comparison of the long-time behaviour of expected returns for Pareto (blue), Uniform (orange) and Gaussian (green) distributions. For the rightward Pareto distribution we choose scale parameter $x_m = 1/4$ and tail index $a = 2$, for the rightward Uniform we choose lower and upper bounds to be $c = 0$ and $d=1$ respectively whereas for the rightward Gaussian we choose $\mu = 1/2$ and ${\sigma = 1/2}$, such that the mean utility for the right steps is $1/2$ in all three cases. Similar distributions are chosen for the left steps with a mean utility of $1$ for all three cases. Simulations are for (a) $10^3$ time-steps and (b) $10^5$ time-steps, each for $10^7$ trajectories.}
  \label{fig:otherdistributions}%
\end{figure}

In passing, we comment more briefly on the behaviour for other utility distributions, again with different parameters for left and right steps. In particular, we consider representative examples of the three different classes already observed in the homogeneous case, for the simplified model. As shown in figure~\ref{fig:otherdistributions}, numerics indicate that for all three classes there exists a peak in the average returns of the agent for finite times. Significantly, we observe that for longer times, the position of the peak on the noise axis appears to move towards zero in the case of bounded distributions (e.g., sampling from uniformly distributed utilities) and towards infinity in the case of power-law tails (e.g., Pareto utilities). We conjecture that only for pure exponential distributions is the position of the peak invariant with time, leading to a robust optimal value of noise. However, for other distributions with exponential tails (e.g., Gaussian utilities\footnote{For the simulations we in fact used a truncated Gaussian so that the simplified model identifies $\hat{U}^+$ with the characteristic largest value $\mu + \sigma \Phi^{-1}\left[ 1 - \frac{\Phi(\mu / \sigma)}{X^+}\right]$ after $X^+$ steps (and analogously for left steps). For numerical efficiency, $\Phi^{-1}$ (the inverse of the cumulative distribution function of the standard
normal distribution) was calculated with the error function approximation in \cite{winitzki}. \nopagebreak}) the peak seems to move only slowly, with an anticipated logarithmic dependence on time. Hence, over reasonable time-scales one can still optimize the ensemble average returns with a finite value of noise. 

Note that throughout this paper we have only considered a dependence on the peak of the experiences of the agent. Based on previous observations of the homogeneous case, we expect that including here a dependence on end-contributions (i.e., the most recent experiences) could effectively change the noise scale in the transition probabilities but would not qualitatively change the conclusions of our analysis. However, further work is needed to investigate the details for different distributions. 

The main result of this paper, namely that for distributions with exponential tails there is a finite value of noise which optimizes the ensemble average returns, means in practice that a certain level of noise is desirable for agents to explore a range of potential options and become trapped in the best one. Of course, an individual agent unaware of the exact utility distributions is not able to explicitly perform this optimization. However, there may be possibilities for strategic interventions on the part of an external organization to optimize returns by controlling the value of noise (e.g., by adding some disruption) or by modifying the utility tails for a certain number of time-steps (e.g., those at the beginning of the walk). This result is only a stepping stone towards more realistic models of behavioural economics; in particular an obvious extension would be to consider the interaction of many agents with different forms of opinion/voter dynamics involving the peak of past experiences (see \cite{polish} for related work in the homogeneous case). There is much scope for both theoretical and practical work in this direction.

\section*{Declaration of competing interest}
The authors declare that they have no known competing financial interests or personal relationships that could have appeared to influence the work reported in this paper.

\section*{Acknowledgments}
This research utilized Queen Mary's Apocrita HPC facility, supported by QMUL Research-IT [http://doi.org/10.5281/zenodo.438045]. R.J.H. gratefully acknowledges an External Fellowship
from the London Mathematical Laboratory.

\appendix

\section*{Appendix}
\label{appendix}
\setcounter{equation}{0}
\renewcommand{\theequation}{A.\arabic{equation}}

Here we explain in more detail the process of obtaining the semi-analytical approximation for the expected utility in the low-noise regime. We note that our simplified model with exponential utilities can be considered as the discrete-time stochastic evolution of two integer populations $A$ and $B$ (corresponding to the number of right and left steps respectively, i.e., $X^{\pm}$) modelled as a generalized P\'olya urn, with transition probabilities
\begin{align}
\label{generalizedpolyaurn1}
    & \text{Pr}\left( \left( A, B \right) \rightarrow  \left( A + 1, B \right) \right) = \frac{A^{\alpha^+}}{A^{\alpha^+} + B^{\alpha^-}}, \\ 
\label{generalizedpolyaurn2}    
    &\text{Pr}\left( \left( A, B \right) \rightarrow  \left( A, B + 1 \right) \right) = \frac{B^{\alpha^-}}{A^{\alpha^+} + B^{\alpha^{-}}},
\end{align}
where $\alpha^{\pm} \in \mathbb{R}$ are exponents, which correspond in our model to $1/ \lambda^{\pm}$. In general for the urn process the initial condition of the populations is given by $(A, B) = (a, b)$, with $a,b>0$. To determine the probability of escaping the trapping state, we are interested in finding out whether the smaller population out of the two will eventually dominate the other. In other words, we want to know if populations that start at state $(a, b)$, with $a \neq b$, cross the ``diagonal'' $A = B$. We thus aim to extend the considerations of \cite{kearney} for the first passage time (i.e., the first time that the system reaches the state $A = B$) and related quantities. To be concrete, \cite{kearney} treats the homogeneous case, $ \alpha^+ = \alpha^-$, whereas we are interested in the heterogeneous case $\alpha^+ \neq  \alpha^-$. Without loss of generality, we choose $\alpha^+ < \alpha^-$ corresponding to $\lambda^+ > \lambda^-$. In contrast to the homogeneous case, we assume that a system with initial conditions $a > b$ does not return to states with $A >B$ after it crosses the line $A = B$. This can be motivated from the potential landscape picture in the low-noise regime (see Fig.~\ref{fig:potentialnoise02}) where crossing the top of the hill towards the bottom valley is typically a one-way route, with the agent not revisiting states up the hill. We expect this to be a good approximation in the limit $T \to 0$. 

To investigate such first passage times, we follow \cite{kearney} and use the method of dynamical embedding \cite{oliveira1, oliveira2, oliveira3}; loosely speaking this means we embed the discrete-time process into a continuous-time process and notice that such a continuous-time process can be then treated as two independent birth processes \cite{krishna, harris, hill, davis, svante, robin, william}. In particular, within this continuous-time picture, the probability there is a jump to the state $A + 1$ from state $A$ in the first birth process before a jump to state $B + 1$ from state $B$ in the second birth process has the same functional form as \eqref{generalizedpolyaurn1}. This implies that the transition-path probabilities within this continuous-time framework are the same as for the original discrete-time model even though the transition times are different. 

The quantity we are interested in calculating is the exit probability $E_n(a, b)$, which is defined as the probability that a given trajectory will cross the diagonal $A = B$ at or before the point $(n,n)$. Given $a > b$, this probability can expressed in terms of the independent birth processes, $L(t)$ for left steps and $R(t)$ for right steps, as the probability $L(t)$ reaches the state $n+1$ before $R(t)$ reaches the state $n+1$. A straightforward adaption of the argument in \cite{kearney}, yields the integral expression
\begin{equation}
    \label{exitprobability}
    E_n(a,b) = \int_{0}^{+ \infty} p_L \left( b, n+1;t \right) \left( 1 - P_R \left( a, n+1;t \right) \right)\text{d}t,
\end{equation} 
where $P_R \left( a, k;t \right)$ is defined as the probability of the right birth process $R(t)$ being in state $k$ at time $t$ starting from $a$, whereas $p_L \left( b, k ;t \right)$ is, correspondingly, the probability density function for the left birth process $L(t)$, to reach the state $ k$ at time $t$, starting from $b$. Note that the equivalent expression in \cite{kearney} has a prefactor of $2$; this is absent in our case because we assume that once the diagonal has been reached the process never returns to states with $A > B$. Following again \cite{kearney} we have
\begin{equation}
    \label{probabilitydensityleft}
    p_L \left( b, k;t \right) = \sum_{\kappa = b}^{k-1}  \kappa^{\alpha^-}\left\{ \prod_{l =  b, l \neq \kappa}^{k-1} \frac{l^{\alpha^-}}{l^{\alpha^-} - \kappa^{\alpha^-}}  \right\}e^{-\kappa^{\alpha^-}t}
\end{equation}
and
\begin{equation}
    \label{probabilityright}
    P_R \left( a, k;t \right) = 1 - \sum_{j = a}^{k-1} \left\{ \prod_{l =  a, l \neq j}^{k-1} \frac{l^{\alpha^+}}{l^{\alpha^+} - j^{\alpha^+}}  \right\}e^{-j^{\alpha^+}t}.
\end{equation}
This leads to an explicit expression for the exit probability\footnote{The calculation makes use of the partial fraction identity
\begin{equation}
    \label{exactexitprob}
    \sum_{j = a}^{k-1} \left\{ \prod_{l = a, l \neq j}^{k-1} \frac{l^{\alpha}}{l^{\alpha} - j^{\alpha}}  \right\} \frac{j^{\alpha}}{j^{\alpha} + s} = \prod_{j = a}^{k-1}\frac{j^{\alpha}}{j^{\alpha} + s}.
\end{equation}}:
\begin{equation}
    \label{exitprobabilityexact}
    E_n(a,b) = \sum_{j = a}^{n} \left\{ \prod_{l = a, l \neq j}^{n} \frac{l^{\alpha^{+}}}{l^{\alpha^{+}} - j^{\alpha^{+}}}  \right\} \left\{  \prod_{\kappa = b}^{n} \frac{\kappa^{\alpha^{-}}}{\kappa^{\alpha^{-}} + j^{\alpha^{+}}} \right \}.
\end{equation}

The exit probability allows us to approximately calculate the chance of the agent becoming trapped: $1 - E_{\infty}(a, b)$ represents the probability the agent remains taking predominantly the less optimal choice, in the long-time limit. [Recall that for \eqref{exitprobabilityexact}, the initial condition $(a, b)$ has $a > b$ and $b>0$.] In fact, assuming that in this limit the system is found in one of the stable fixed points, we can express the average expected returns as
\begin{equation}
    \label{averagereturnsapproxappendix}
    \langle U \rangle = \alpha^+ \times p^+ +  \alpha^- \times \left( 1 - p^+ \right),
\end{equation}
where $\alpha^+ = 1/ \lambda^+, \alpha^- = 1/ \lambda^-$ and $p^+$ is the probability of being trapped in the less optimal choice, in the long-time limit, starting at $(0,0)$. Our chief task is thus to express $p^+$ in terms of $E_{\infty} \left( a, b \right)$.

As a first basic approximation we look for all the paths that start at $(0,0)$ and never actually reach the diagonal $X^+ = X^-$. We can extend the definition of the urn process \eqref{generalizedpolyaurn1}/\eqref{generalizedpolyaurn2} to match the initial condition of the simplified model by replacing $A^{{\alpha}^+}$ by $1$ when $A = 0$ and similarly $B^{\alpha^-}$ by $1$ when $B = 0$. The system hence makes the first and second step to the right with probability $1/2 \times 1/2$. The trajectory which makes the first two steps to the right and the third to the left, but does not cross the diagonal, happens with probability
\begin{equation}
\frac{1}{2} \times \frac{1}{2} \times \frac{1}{1 + 2^{\alpha^+}}\times \left(1 - E_{\infty} \left( 2,1  \right) \right). 
\end{equation}
Similarly, the trajectory that makes the first three steps to the right and the fourth to the left, but does not cross the diagonal happens with probability
\begin{equation}
\frac{1}{2} \times \frac{1}{2} \times \frac{2}{1 + 2^{\alpha^+}} \times \frac{1}{1 + 3^{\alpha^+}}\times \left(1 - E_{\infty} \left( 2,1  \right) \right).
\end{equation}
Extending this logic, the probability that the system takes at least two steps to the right at the beginning of the walk and does not cross the diagonal, i.e., remains trapped, is given by the sum
\begin{equation}
    \frac{1}{2} \sum_{m = 2}^{\infty} \left\{ \prod_{q = 1}^{m-1}  \frac{q^{\alpha^+}}{1 + q^{\alpha^+}} \right\} \frac{1}{1 + m^{\alpha^+}} \left( 1 - E_{\infty} \left( m,1  \right) \right).
\end{equation}
A numerical calculation of the infinite sum is costly. However, we can make two further approximations for practical purposes. Firstly, for finite $m$, we replace $E_{\infty}(m,1)$ with $E_{N}(m,1)$, where $N$ is a suitably large constant. Secondly, we assume that for $m > M$ (where $M$ is again a suitably large constant) the probability of the system crossing the diagonal is zero, i.e., $E_{N}(m, 1) =  0$. Hence, the previous probability can now be written as
\begin{equation}
    \frac{1}{2} \sum_{m = 2}^{M} \left\{ \prod_{q = 1}^{m-1}  \frac{q^{\alpha^+}}{1 + q^{\alpha^+}} \right\} \frac{1 - E_{N} \left( m,1  \right)}{1 + m^{\alpha^+}} + \frac{1}{2} \sum_{m = M + 1}^{\infty} \left\{ \prod_{q = 1}^{m-1}  \frac{q^{\alpha^+}}{1 + q^{\alpha^+}} \right\} \frac{1}{1 + m^{\alpha^+}}.
\end{equation}
Finally, we note that the remaining infinite sum has to equal the probability of starting with at least two right steps, which is $1/4$, minus the probability of all the paths included in the sum from $m = 2$ to $M$:
\begin{equation}
    \frac{1}{2} \sum_{m = M + 1}^{\infty} \left\{ \prod_{q = 1}^{m-1}  \frac{q^{\alpha^+}}{1 + q^{\alpha^+}} \right\} \frac{1}{1 + m^{\alpha^+}} = \frac{1}{4} - \frac{1}{2} \sum_{m = 2}^{M} \left\{ \prod_{q = 1}^{m-1}  \frac{q^{\alpha^+}}{1 + q^{\alpha^+}} \right\} \frac{1}{1 + m^{\alpha^+}}.
\end{equation}
Combining the above expressions yields the first basic approximation for $p^+$:
\begin{equation}
    \label{basicapproximation}
    p^+ \approx
    \frac{1}{4}- \frac{1}{2} \sum_{m = 2}^{M} \left\{ \prod_{q = 1}^{m-1}  \frac{q^{\alpha^+}}{1 + q^{\alpha^+}} \right\} \frac{E_{N} \left( m,1  \right)}{1 + m^{\alpha^+}}.
\end{equation}

Recall that in the above analysis we assumed throughout that all paths that reach the diagonal $X^+ = X^-$ ultimately end up with $X^- > X^+$. The largest error in this approximation is expected to be associated with paths that reach the diagonal at early times. The first correction to \eqref{basicapproximation} thus corresponds to paths that cross the diagonal at $(1,1)$. The probability of paths that pass through $(1,1)$ to $(2, 1)$ and then remain above the diagonal is
\begin{equation}
    \frac{1}{4} \times \left( 1 - E_{N} \left( 2,1  \right)  \right).
\end{equation}
Taking this into account gives an improved approximation for $p^+$:
\begin{equation}
        p^+ \approx
    \frac{1}{2} \left( 1 - \frac{E_{N} \left( 2,1  \right)}{2} - \sum_{m = 2}^{M} \left\{ \prod_{q = 1}^{m-1}  \frac{q^{\alpha^+}}{1 + q^{\alpha^+}} \right\} \frac{E_{N} \left( m,1  \right)}{1 + m^{\alpha^+}}  \right).
\end{equation}
Substituting this in \eqref{averagereturnsapproxappendix} gives the theoretical $\langle U \rangle$ shown in Fig.~\ref{fig:approximationlownoise}. It can be seen that this is a very good approximation in the small-noise limit but further terms would presumably be needed to accurately capture the behaviour for larger values of noise. Other approximation schemes seem to be possible for values of noise near $1 / \lambda^+$ but are not discussed here.

\bibliography{References}

\end{document}